\documentclass[11pt]{article} 

\usepackage{jcappub}

\usepackage[utf8]{inputenc}   
\usepackage{amsmath}          
\usepackage{amssymb}          
\usepackage{graphicx}         
\usepackage{hyperref}         
\usepackage{geometry}         
\usepackage{siunitx}          
\usepackage{float}            
\usepackage{enumitem}         
\usepackage{tocloft}
\usepackage{tikz}
\usepackage{tikz-3dplot}
\usetikzlibrary{shapes.geometric, arrows, backgrounds}
\usepackage{empheq}
\usepackage{booktabs}
\usepackage{multirow} 
\usepackage{booktabs}
\usepackage[table]{xcolor}

\usepackage{listings}
\usepackage{xcolor}
\usepackage{pgfmath}
\newcommand{\greenfactor}[1]{%
    \pgfmathsetmacro{\value}{min(100,round(100*(#1-1)))}%
    \textcolor{green!80!black!\value!black}{#1}%
}

\usepackage[backend=bibtex, sorting=none]{biblatex}
\DeclareRobustCommand{\d}{\ifmmode\text{d}\else d\fi}
\DeclareRobustCommand{\Cov}{\ifmmode\text{Cov}\else d\fi}
\DeclareRobustCommand{\CMB}{\ifmmode\text{CMB}\else d\fi}
\DeclareRobustCommand{\gal}{\ifmmode\text{gal}\else d\fi}

\newcommand{\br}[1]{\ensuremath{\left( #1 \right)}}
\newcommand{\sbr}[1]{\ensuremath{\left[ #1 \right]}}

\title{Future Parameter Constraints from Weak Lensing CMB and Galaxy Lensing Power- and Bispectra}

\author[1]{Jonas Frugte}
\author[2]{P. Daniel Meerburg}

\affiliation[1]{Institute for Theoretical Physics, Utrecht University, Princetonplein 5, 3584 CC Utrecht, The Netherlands}
\affiliation[2]{Van Swinderen Institute for Particle Physics and Gravity, University of Groningen, Nijenborgh 4, 9747 AG Groningen, The Netherlands}

\emailAdd{j.s.a.frugte@uu.nl}
\emailAdd{p.d.meerburg@rug.nl}
\abstract{
Upcoming stage 4 surveys, such as the Simons Observatory, LSST, and Euclid, are poised to measure weak gravitational lensing of the Cosmic Microwave Background (CMB) and galaxies with unprecedented precision. While the power spectrum is the standard statistic used to analyze weak lensing data, non-Gaussianity from non-linear structure growth encodes additional cosmological information in higher-order statistics. We forecast the ability of future surveys to constrain cosmological parameters using the weak lensing power spectrum and bispectrum from both CMB and galaxy surveys, including their cross-correlations. We consider an eight-parameter model ($\Lambda$CDM + $\sum m_\nu$ + $w_0$) and assess constraints for stage 4 survey specifications. In the absence of systematics, both the CMB and galaxy lensing bispectra are found to be detectable at high signal-to-noise. We test two priors: a ``strong" one based on constraints from CMB temperature and $E$-mode polarization anisotropies, and a ``weak" one with minimal assumptions. With the weak prior, the bispectrum significantly improves parameter constraints by breaking degeneracies. For strong priors, improvements are more limited, especially for the CMB bispectrum. On small scales, where non-linear effects dominate, the bispectrum’s constraining power can rival that of the power spectrum. We also find strong synergy between CMB and galaxy lensing; combining both probes leads to tighter constraints, particularly on neutrino mass. It was recently found that the CMB lensing bispectrum is strongly affected by the Born approximation, so we also consider post-Born corrections but find that our main conclusions remain the same. These results highlight the potential of higher-order lensing statistics and motivate further work on neglected effects such as non-Gaussian covariance, instrumental systematics, and baryonic feedback.
}

\begin{document}

\maketitle





\section{Introduction}

Weak lensing, first detected in 2007 for radiation from the CMB \cite{Smith2007} and in 2000 for that of galaxies \cite{Bacon2000, Kaiser2000, Wittman_2000, Waerbeke2000}, has been used as a probe to constrain cosmological parameters for over a decade \cite{kilbinger2015cosmology, Waerbeke2000, Bartelmann2001, Planck2018Lensing}. A range of upcoming surveys, such as the Simons Observatory (SO) \cite{Ade2019}, the Legacy Survey of Space and Time (LSST) \cite{Ivezic2019}, and \textit{Euclid} \cite{Laureijs2011} aims to measure weak lensing of the cosmic microwave background (CMB) and galaxies respectively. These upcoming surveys are expected to achieve significantly higher accuracy than their predecessors, leading to tighter constraints on cosmological parameters and advancing our overall understanding of cosmology. Typically, the standard summary statistic of interest is the power spectrum (or equivalently, the two-point function) of the lensing potential (as well as cross-correlations with galaxy clustering). This can be measured in terms of the lensing convergence $\kappa$, lensing shear $\gamma$, or lensing potential $\psi$. In the weak lensing regime and under the Born approximation, they are all equivalent, in the sense that they can be directly converted into one another (appendix \ref{sec:weaklensstats}). In CMB surveys, the temperature and polarization anisotropies are used to reconstruct the lensing potential \cite{Hu_2002}, while with galaxies, one instead calculates lensing effects by measuring the lensing shear, which can be deduced from the ellipticity distribution of the observed galaxies \cite{HoekstraJain2008}.

Even with Gaussian initial conditions, there are significant amounts of non-Gaussianity in the weak lensing signal, especially in galaxy lensing due to the fact that lensing indirectly traces the more recent distribution of density \cite{Bernardeau1997,Takada2003}. Higher order spectra, such as the bispectrum, have already been shown to be useful for constraining cosmological parameters and breaking parameter degeneracies. A first detection of non-Gaussian clustering using lensing has recently been achieved by cross correlating the matter overdensity squared with CMB lensing convergence \cite{Farren:2023yna, Harscouet:2025ksm}. In addition to looking at power spectra, a natural next step is to measure weak lensing bispectra. Both the CMB and galaxy lensing power spectra are detectable with current surveys \cite{Das2011, Planck2013Lensing, Planck2015Lensing, Planck2018Lensing}. On the other hand, while the bispectrum has been measured for galaxy lensing signals \cite{Simon_2013}, a detection of the CMB lensing bispectrum is still pending \cite{Kalaja_2023, Namikawa_2016}. Upcoming surveys, such as the ones mentioned earlier, may be able to detect this higher order spectrum. Therefore it is timely to consider the potential of higher-order spectra to constrain cosmological parameters.

In our analysis, we will include post-Born corrections. The CMB and galaxy lensing power spectra are typically affected by less than $1\%$\footnote{These corrections should still be taken into account when fitting parameters to data in future surveys to avoid significant biasing, see for example ref. \cite{Marozzi_2017}} \cite{postborn_pratten_lewis, galpostbornlpscorr, Marozzi_2016, Marozzi_2018}. The galaxy lensing bispectrum is only affected by a few percent at most \cite{galpostbornlbscorr}. The only significant effect is on the CMB lensing bispectrum, as was shown in ref. \cite{postborn_pratten_lewis}. In particular, for the folded configuration the effect can be of the same magnitude as the lensing bispectrum due to nonlinear large scale structure \eqref{eq:lensingbispectrum}. Including post-Born effects  will allow us to determine how future parameter constraints are affected by more accurate modeling.

In this paper, we aim to answer these questions by adopting experimental configurations similar to those of next-generation surveys. We are especially interested in seeing if approximate parameter degeneracies can be broken by combining CMB and galaxy weak lensing power- and bi-spectra. 
We will consider $\Lambda$CDM parameters and extensions thereof. Of particular interest are the sum of neutrino masses $\sum m_\nu$ and dark energy equation of state parameter, $w_0$.


The structure of this paper is as follows: we introduce relevant formulas for weak lensing statistics, nonlinear matter bispectrum, and Fisher matrix formalism in section \ref{sec:theory}. We specify details such as the fiducial values of the parameters and noise models used in section \ref{sec:calcdetails}. Parameter constraints and signal-to-noise ratios are presented in section \ref{sec:results} and discussed in section \ref{sec:discussion}. Useful derivations relevant to this paper can be found in the appendices.

All code used to derive the results in this paper is publicly available on GitHub at \url{https://github.com/Jonas-Frugte/fisher_calc_weak_lensing}.


\section{Background}
\label{sec:theory}
\subsection{Weak lensing spectra}
Radiation from cosmological objects is distorted due to gravitational lensing. Due to the low density of the cosmic web, the average deflection of a photon as it propagates through the universe is relatively weak. We thus work under the assumption that all deflection angles are small, this is referred to as the weak lensing regime. Weak lensing is quantified through the deflection field $\mathbf d(\hat {\mathbf n})$ which equals the difference between the observed angle of a point in the sky and the true (unlensed) angle. This field is the gradient of the lensing potential, $\psi(\hat {\mathbf n})$, which is expressed as a weighted integral of gravitational potential along the line-of-sight from the observer to the source.

In the case of CMB surveys, lensing alters the statistical properties of the temperature and polarization fields and can thus be calculated by comparing the observed signal to the expected unlensed signal (see e.g. \cite{Maniyar2021_QE_CMBLensing, Hirata_2003}). In galaxy surveys, lensing alters the ellipticities of observed galaxies. If a large enough number of galaxies are observed, this effect can be separated from the intrinsic ellipticities of the galaxies which allows us to estimate the lensing potential.

To constrain cosmological parameters we can then look at the lensing potential of the CMB, $\psi_\CMB$, and of galaxy surveys, $\psi_\gal$. These are directly related to the matter power and bispectra as
\begin{align}
    C^{\psi_X\psi_Y}_\ell
    & = \frac{9}{\ell^4}\Omega_m^2H_0^4\int_0^{\chi_*} \chi^2 \d\chi a(\eta_0-\chi)^{-2}W_X(\chi)W_Y(\chi)P^\delta(\ell/\chi,\eta_0-\chi), \label{eq:lensingpower spectrum}\\
    B^{\psi_X\psi_Y\psi_Z}_{\ell_1\ell_2\ell_3} &= \sqrt{\frac{(2\ell_1 + 1)(2\ell_2 + 1)(2\ell_3 + 1)}{4\pi}} \begin{pmatrix} \ell_1 & \ell_2 & \ell_3 \\ 0 & 0 & 0 \end{pmatrix} \frac{27}{\ell_1^2\ell_2^2\ell_3^2}\Omega_m^3H_0^6 \notag\\
    & \quad \times \int \chi^2\d \chi a(\eta_0-\chi)^{-3}W_X(\chi)W_Y(\chi)W_Z(\chi)  B^\delta(\{\ell_i/\chi\}, \eta_0-\chi), \label{eq:lensingbispectrum}
\end{align}
with $X, Y, Z \in \{\CMB, \gal\}$. A derivation can be found in appendices \ref{sec:weaklensing} and \ref{sec:weaklensstats}, or in the literature; see, e.g. \cite{Bartelmann2001}. Here, $\Omega_m$ is the present day matter density parameter, $H_0$ is the present-day Hubble constant, $a(\eta)$ is the scale factor, $\eta_0$ is the conformal time today and $\chi$ is the comoving radial distance. $\chi_*$ represents the distance to surface of last scattering. $P^\delta(k, \eta)$ is the matter power spectrum. $B^\delta(k_1, k_2, k_3, \eta)$ is the matter bispectrum. $W_X(\chi)$ is the window function. Finally,
$$
\begin{pmatrix}
    \ell_1&\ell_2&\ell_3 \\ m_1 & m_2 & m_3
\end{pmatrix},
$$ 
is the Wigner 3-j symbol. The difference between CMB and galaxy lensing is the window function, defined as
\begin{equation}
    W_X(\chi) = \int_\chi^{\infty}\d\chi' p_X(\chi')\frac{\chi'-\chi}{\chi'\chi},
    \label{eq:windowfuncdef}
\end{equation}
with $p_X$ the radial distribution of the radiation source. For the CMB we take $p(\chi) = \delta(\chi - \chi_*)$. The window function represents the distribution of redshifts at which a CMB photon or galaxy photon is deflected. As a result, the CMB lensing window function is broader and peaks at higher redshift compared to the galaxy window function.

Our convention for defining the linear and non-linear matter power spectrum is
\begin{equation*}
    \langle \delta(\mathbf k, \eta) \delta(\mathbf{k}',\eta) \rangle = (2\pi)^3\delta_D(\mathbf k + \mathbf k ')P^\delta(\mathbf k, \eta),
\end{equation*}
and for the bispectrum it is 
\begin{equation*}
    \langle \delta(\mathbf k_1, \eta) \delta(\mathbf k_2, \eta)\delta(\mathbf k_3, \eta) \rangle = (2\pi)^3\delta_D(\mathbf k_1 + \mathbf k_2 + \mathbf k_3)B^\delta(k_1,k_2,k_3, \eta),
\end{equation*}
with $\delta$ the fractional matter density perturbation field and $\delta_D$ the Dirac delta function.

\subsection{Nonlinear matter bispectrum}

An approximate form of nonlinear matter power spectrum can be obtained using numerical codes such as CAMB \cite{Lewis2000}. The nonlinear matter bispectrum was calculated from the power spectrum using a fitting formula based on perturbation theory in \cite{bispfit}. It is given by
\begin{equation}
    B^\delta(k_1, k_2, k_3, \chi) = 2 F_2(k_1, k_2, z) P^\delta(k_1, z) P^\delta(k_2, z) + \text{2 perm}, \label{eq:matterbispectrum}
\end{equation}
where \( P^\delta \) is the nonlinear matter power spectrum\footnote{Compare to the tree level bispectrum where we instead use the linear power spectrum.}, and the kernel \( F_2 \) is modified from the tree level result with factors \( a(k, z) \), \( b(k, z) \), and \( c(k, z) \):
\begin{equation}
    F_2(k_1, k_2, z) = \frac{5}{7} a(k_1, z) a(k_2, z) + \frac{k_1^2 + k_2^2}{2 k_1 k_2} b(k_1, z) b(k_2, z) \cos \theta + \frac{2}{7} c(k_1, z) c(k_2, z) \cos^2 \theta.
\end{equation}
They are defined as:
\begin{equation}
    a(k, z) = \frac{1 + \sigma_8^{a_6}(z) \sqrt{0.7} Q(n_{\text{eff}}) (q^{a_1})^{n_{\text{eff}} + a_2}}{1 + (q^{a_1})^{n_{\text{eff}} + a_2}},
\end{equation}
\begin{equation}
    b(k, z) = \frac{1 + 0.2 a_3 (n_{\text{eff}} + 3) (q^{a_7})^{n_{\text{eff}} + 3 + a_8}}{1 + (q^{a_5})^{n_{\text{eff}} + 3.5 + a_8}},
\end{equation}
\begin{equation}
    c(k, z) = \frac{1 + \left[ \frac{4.5 a_4}{1.5 + (n_{\text{eff}} + 3)^4} \right] (q^{a_5})^{n_{\text{eff}} + 3 + a_9}}{1 + (q^{a_5})^{n_{\text{eff}} + 3.5 + a_9}}.
\end{equation}
Here, \( Q(n_{\text{eff}}) \) is given by:
\begin{equation}
    Q(x) = \frac{4 - 2x}{1 + 2^{x+1}}.
\end{equation}
The effective spectral index of the linear power spectrum is defined as:
\begin{equation}
    n_{\text{eff}} \equiv \frac{d \ln P^\delta_{\text{lin}}(k)}{d \ln k}.
\end{equation}
Note that we ``smooth'' $n_{\text{eff}}$ to avoid effects due to BAO, as described in \cite{bispfit}.
Additionally, \( q \) is given by:
\begin{equation}
    q = \frac{k}{k_{\text{NL}}},
\end{equation}
where \( k_{\text{NL}} \) is the scale at which nonlinearities become significant, satisfying:
\begin{equation}
    4 \pi k_{\text{NL}}^3 P^\delta_{\text{lin}}(k_{\text{NL}}, 0) = 1.
\end{equation}
The coefficients \( a_i \) are:
\begin{align*}
    a_1 &= 0.484, \quad a_2 = 3.740, \quad a_3 = -0.849, \quad a_4 = 0.392, \\
    a_5 &= 1.013, \quad a_6 = -0.575, \quad a_7 = 0.128, \quad a_8 = -0.722, \quad a_9 = -0.926.
\end{align*}

\subsection{Post-Born corrections}
One of the approximations made in the derivation of the weak lensing power- and bispectra (equations \eqref{eq:lensingpower spectrum} and \eqref{eq:lensingbispectrum}) is that the integral is performed along the line of sight to the light source. To get an analytically correct result, one would instead need to solve a differential equation with a feedback loop where the effects of gravitational lensing affect the trajectory and shape of given bundle of light rays, which in turn changes the way future gravitational lenses affect the bundle. One can also choose to make order by order corrections to the Born approximation. As outlined in the introduction there is motivation to understand the effects of these corrections, so we present them here.

Under the Born approximation, the elements of the distortion tensor, $\psi_{ab}$, (from which shear, convergence, and rotation are constructed) equal second derivatives of the lensing potential. We thus calculate the potential instead of the distortion tensor itself. When post-Born effects are taken into account this is no longer true and we are required to work with $\psi_{ab}$ directly. Following \cite{postborn_pratten_lewis}, the leading order post-Born correction to the distortion tensor for a point source at distance $\chi_s$ is given by
\begin{equation}
    \left( \psi_{ab}^{\text{p.s.}}  \right)^{(2)}(\chi_s) := 2 \int^{\infty}_0 d \chi^{} \, \chi^2 \, \frac{\chi_s - \chi}{\chi_s\chi}\Theta(\chi_s - \chi) \, \left[ - \Psi_{,ac} (\chi) \left(\psi^{\text{p.s.}}_{cb}\right)^{(1)} (\chi) + \Psi_{,acd} (\chi) \, \delta x_d^{(1)} (\chi^{}) \right].
    \label{eq:pbcorrectionpointsorce}
\end{equation}
Here $\left(\psi_{cb}^{(p.s.)}\right)^{(1)}(\chi)$ is the deformation for a point source at $\chi$ under the Born approximation as defined earlier. $\delta x_d^{(1)}(\chi)$ is the first order correction to the location of the lensed light ray at a distance $\chi$, given by
\begin{equation}
\delta x_a^{(1)} = -2 \int^{\chi}_0 d \chi^{\prime} \, \frac{\chi-\chi'}{\chi\chi'} \, \chi \chi' \, \Psi_{,a} (\chi').
\end{equation}
To adapt the above to a source with radial distribution $p(\chi)$ (which in our case will be a population of galaxies) we integrate with respect to $\chi_s$ and over the result from \eqref{eq:pbcorrectionpointsorce} as follows:
\begin{equation}
    (\psi_{ab}^{\text{gal}})^{(i)} = \int_0^\infty d \chi_s p(\chi_s)\left(\psi_{ab}^{(p.s.)}\right)^{(i)}(\chi_s).
\end{equation}
This step essentially just changes $(\chi_s - \chi)/(\chi_s\chi)\Theta(\chi_s-\chi)$ in equation \eqref{eq:pbcorrectionpointsorce} to the window function $W(\chi)$ defined in equation \eqref{eq:windowfuncdef}.

The leading order correction to the convergence bispectrum of sources 1, 2, 3 is given as
\begin{equation}
    B^{\kappa_{(1)}\kappa_{(2)}\kappa_{(3)}}_{L_1 L_2 L_3} = 2\frac{\mathbf L_1\cdot\mathbf L_2}{L_1^2L_2^2}\left[ \mathbf L_1\cdot \mathbf L_3 \mathcal M_s^{(1)(2)(3)}(L_1,L_2)
 +\mathbf L_2\cdot \mathbf L_3 \mathcal M_s^{(2)(1)(3)}(L_2,L_1)\right] + \text{cyc. perm.},
\end{equation}
where
\begin{equation}
    \mathcal M_s^{(1)(2)(3)}(L_1,L_2) := L_1^4\int_0^{\chi_s}d\chi \frac{W^{(1)}(\chi)W^{(3)}(\chi)}{\chi^2} P_{\Psi\Psi}\left(\frac{L_1}{\chi}, z(\chi)\right) C_{L_2}^{\kappa_{(2)}\kappa_{(3)}}(\chi,\chi_s),
\end{equation}
and
\begin{equation}
    C_{L_2}^{\kappa_{(2)}\kappa_{(3)}}(\chi,\chi_s) := \int_0^{\infty}d \chi' W^{(2)}(\chi')\frac{\chi_s-\chi}{\chi_s\chi}\Theta(\chi_s-\chi)P_{\Psi\Psi}(\frac{L_2}{\chi'},\chi').
\end{equation}

\subsection{Redshift tomography}
\begin{figure}
    \centering
    \includegraphics[width=\textwidth]{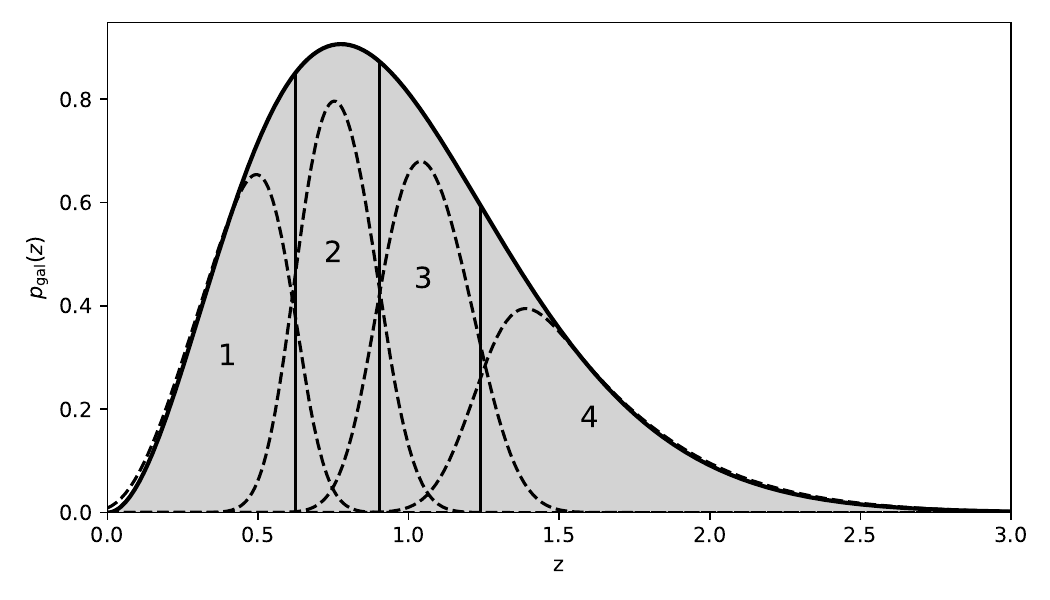}
    \caption{Our galaxy redshift distribution (\ref{eq:galzdist}) split up into 4 bins. The dashed lines correspond to the actual distributions of the bins after taking into account galaxy redshift uncertainty. We assumed no galaxy bias in the redshift estimation and a gaussian error given by $\sigma_z = 0.05 (1+z)$. This is representative of Euclid level of redshift uncertainty \cite{EuclidSciRD}.}
    \label{fig:galaxyredshiftbinning}
\end{figure}

More information can be extracted from galaxy weak lensing surveys by binning the galaxy population \cite{Hu_1999}, see figure \ref{fig:galaxyredshiftbinning}. This gives additional lensing power spectra and bispectra. We get auto spectra such as $C^{11}_{\ell}$ and cross spectra such as $C^{12}_{\ell}$. The same is true for the bispectra. The corresponding noise spectra are larger because the population in each bin is of course smaller. Nevertheless, the overall constraining power is usually significantly improved by using such ``redshift tomography'' (see e.g. \cite{Hu_1999}). The constraining power is especially improved if the different bins are lensed differently, i.e. if the cross-bin spectra are small. One can also take into account the significant uncertainty in galaxy redshift in a photometric survey. This is done by convolving the galaxy distribution function with the distribution function of a galaxy's true redshift given its observed redshift. See again figure \ref{fig:galaxyredshiftbinning}.

The obvious question when analyzing (simulated) data is how many bins to use and where their boundaries should be. There is no obvious answer for this and different options should be tested for different bin configurations. Using more bins generally leads to increasingly tight constraints but with diminishing returns. For example, in \cite{Duncan_2013} they find little improvement in constraints from galaxy shear power spectra when going beyond 4 bins. Some considerations are:
\begin{enumerate}
    \item If the noise starts to be bigger than the signal for each bin, further improvements in constraints will be minimal \cite{Hu_1999}.
    \item The bigger the redshift uncertainty, the more overlap between bins, the more their signals are correlated. With higher redshift uncertainty it thus pays less to have a larger number of bins.
    \item In the case of this paper we compute Fisher matrices also taking into account cross and auto bispectra, leading to a large amount of terms that need to be calculated and summed over. Using a large amount of bins will further extend this computational time. It is thus practical to stay on the lower end with the amount of bins used.
\end{enumerate}

\subsection{Fisher matrix analysis}
The signal-to-noise (SNR) and parameter constraints are obtained through a Fisher matrix analysis as is standard in cosmology \cite{dodelson2020modern, Knox1995, Tegmark1997}. A full derivation of the equations used below can be found in appendix \ref{sec:fisher}. The auto- and cross- power spectrum Fisher matrix is given as
\begin{equation*}
    F_{\alpha\beta} = \sum_{\ell} \sum_{XY}\sum_{X'Y'}(2\ell+1)\partial_\alpha C_\ell^{XY} (C^{-1})_\ell^{XX'}(C^{-1})_\ell^{YY'} \partial_\beta C^{X'Y'}_\ell,
\end{equation*}
where $X$, $Y$, etc. are iterated over the tracers that we correlate\footnote{To be clear, for CMB lensing the tracer is just CMB lensing. For galaxy lensing the tracers are the different redshift bins (so 4). For CMB + galaxy lensing the tracers are CMB lensing plus the redshift bins (so 5)}. The powerspectra matrices are defined as
\begin{equation*}
    C_{\ell} := \begin{pmatrix}
        C^{XY}_\ell
    \end{pmatrix}.
\end{equation*}
For example, when correlating all data, we get
\begin{equation*}
    C_{\ell} := \begin{pmatrix}
        C^{\psi_{\text{CMB}}\psi_{\text{CMB}}}_\ell & C^{\psi_{\text{CMB}}\psi_{\text{gal bin 1}}}_\ell & \cdots \\
        C^{\psi_{\text{CMB}}\psi_{\text{gal bin 1}}}_\ell & C^{\psi_{\text{gal bin 1}}\psi_{\text{gal bin 1}}}_\ell & \cdots \\
        \vdots & \vdots & \ddots
    \end{pmatrix}.
\end{equation*}

For the bispectra we instead have 
\begin{equation*}
    F_{\alpha\beta} = \sum_{\ell_1 \leq \ell_2 \leq \ell_3} \frac{\mathcal P _{\ell_1\ell_2\ell_3}}{6}
    \sum_{XYZ}\sum_{X'Y'Z'} 
    \partial_\alpha B^{X Y Z}_{\ell_1 \ell_2 \ell_3} 
    (C^{-1})^{X X'}_{\ell_1}
    (C^{-1})^{Y Y'}_{\ell_2}
    (C^{-1})^{Z Z'}_{\ell_3}
    \partial_\beta B^{X' Y' Z'}_{\ell_1 \ell_2 \ell_3},
\end{equation*}
with $\mathcal P_{\ell_1\ell_2\ell_3}$ defined as the number of distinct permutations that can be made with $\ell_1\ell_2\ell_3$. When only considering auto spectra the formulas simplify to:
\begin{equation*}
    F_{\alpha\beta} = \sum_{\ell} \sum_{XY}\sum_{X'Y'}\frac{2\ell+1}{2}\frac{\partial_{\alpha}C_\ell^{XX} \partial_\beta C^{XX}_\ell}{(C_\ell^{XX})^2},
\end{equation*}
and (for bispectra, assuming Gaussian, diagonal, covariance)
\begin{equation*}
    F_{\alpha\beta} = \sum_{\ell_1 \leq \ell_2 \leq \ell_3} \frac{\mathcal P _{\ell_1\ell_2\ell_3}}{6}
    \frac{\partial_\alpha B^{X X X}_{\ell_1 \ell_2 \ell_3} 
    \partial_\beta B^{XXX}_{\ell_1 \ell_2 \ell_3}}{C^{XX}_{\ell_1}C^{XX}_{\ell_2}C^{XX}_{\ell_3}}.
\end{equation*}
If only some fraction of the sky, $f_{\text{sky}}$, is measured, all Fisher matrices are multiplied by $f_{\text{sky}}$ \cite{Takada2003}.

\section{Experimental parameters and priors}\label{sec:calcdetails}
\subsection{Fiducial cosmology}
Our fiducial cosmology is based on the \textit{Planck} results \cite{planckresults} (see table \ref{tab:fiducialpars}). We choose to constrain the standard parameters of the $\Lambda$CDM model as well as neutrino mass, $m_\nu$, and dark energy equation of state parameter, $w_0$. The reason to include these extra degrees of freedom is to explore the potential of future surveys on these extensions of $\Lambda$CDM. For example, the DESI collaboration \cite{Roy2024DESI} recently showed that measurements of baryon acoustic oscillations give a 2.5 - 4 $\sigma$ tension with a non-evolving dark energy equation of state model, so it is worth considering if future weak lensing experiments could tell us more about this tension and if higher order statistics would benefit the constraints on these tensions. Regarding neutrinos, depending on how tightly stage 4 surveys will be able to constrain neutrino mass, the results could tell us more about their mass hierarchy and serve as a test for neutrino mass constraints obtained from other probes, such as baryon acoustic oscillations. 

Finally, we also need to take into account baryonic feedback. It has been shown that not taking into account baryonic feedback will lead to a bias of multiple sigma in, for example, $S_8$, $\Omega_m$, $\Omega_b$ when analyzing 2 and 3 point correlations of Euclid mock data \cite{Burger_2026}. We model the effects by using HMCODE2020 \cite{Mead_2021} and marginalizing over $\log T_{\text{AGN}}$ in addition to the parameters described above. It was shown in \cite{Mead_2021} that this one parameter allows models of the nonlinear power spectrum to incorporate baryonic feedback at a high level of accuracy (compared to models with more degrees of freedom for baryonic feedback). In this way we are able to make our forecast considerably more realistic with minimal extra computational cost. It is worth noting that the baryonic feedback in HMCODE2020 was modeled only in such a way as to make the matter power spectrum align with results from simulations. We are assuming that using the adjusted matter power spectrum in the fitting formula for the matter bispectrum, eq. \ref{eq:matterbispectrum}, also adjusts the matter bispectrum accurately. We believe this is a reasonable assumption given the limited knowledge on the subject in the literature and baryonic feedback being a ``second order'' effect. Nevertheless, this should be further investigated in future work.

\subsection{Choice of priors}
Throughout this paper we consider two priors. The first we refer to as our ``weak prior'' and is based on the one used in \cite{Planck2018Lensing} (see table \ref{tab:fiducialpars}). This prior only restricts $\Omega_b h^2$ and $n_s$ significantly, as those are not well constrained by lensing spectra alone. Our differences with \cite{Planck2018Lensing} are that (i) we do not fix $\tau$, but instead take it to have a SNR of 1 and added a weak $A_s$ prior as well, (ii) instead of a flat prior for the Hubble constant ($40 < H_0 < 100$) we use a Gaussian distribution with the same standard deviation, (iii) we added a weak constraint on $\Omega_c h^2$, and (iv) we also vary neutrino mass and $w_0$ but do not assume any constraints on them in our priors. The purpose of the weak prior is as follows. The weak lensing power spectra, in particular, exhibit complete insensitivity to certain parameters. Consequently, relying solely on them to constrain the entire cosmological model results in very poor constraints. Future lensing surveys will of course use information from surveys measuring other statistics such as those derived from galaxy clustering and the primary CMB anisotropies. By adding a prior based on these (primary) statistics our results become more representative of what we should expect is possible. Additionally, by considering a conservative prior, it remains clear how the different lensing spectra complement each other. The other prior we consider is based on the CMB temperature and $E$-mode polarization cross- and auto-power spectra from $\ell=30$ to $\ell = 2000$ assuming the same noise properties as used for the CMB weak lensing reconstruction noise. Adopting these priors  serves to test how weak lensing statistics are able to further tighten constraints beyond measuring primary statistics. It is worth noting that the constraint for $\tau$ and $w_0$ are tighter than in other literature (see \cite{cmbs4sciencebook} and \cite{Namikawa_2016} for comparable $\tau$ and $w_0$ constraints, respectively). Based on a number of internal tests\footnote{
This included looking at the accuracy of numerical derivatives w.r.t. cosmological parameters, testing the numerical stability of Fisher matrices w.r.t. matrix inversion, and varying our values for detector noise and beam width.
} we suspect that this is partly due to us not taking into account foregrounds in our noise models.


\begin{table}[!]
    \centering
    \scriptsize
    \begin{tabular}{|c|c|c|c|c|}
        \hline
        \textbf{parameter} & \textbf{notation} & \textbf{fiducial value} & \textbf{$\sigma$ of weak prior} & \textbf{$\sigma$ of $T + E$ prior}\\
        \hline
        Hubble constant & $H_0$ & $67.4 \, \text{km/s/Mpc}$ & 17.3 & 1.4\\
        \hline
        Physical baryon density parameter & $\Omega_b h^2$ & $0.0223$ & 0.0005 & 0.000065\\
        \hline
        Physical cold dark matter density parameter & $\Omega_c h^2$ & $0.119$ & $0.288$ & 0.00087\\
        \hline
        Scalar spectral index & $n_s$ & 0.965 & 0.02 & 0.0026 \\
        \hline
        Reionization depth & $\tau$ & 0.063 & 0.063 & 0.021 \\
        \hline
        Amplitude of primordial scalar fluctuations & $A_s$ & $2.13 \times 10^{-9}$ & $1\times 10^{-9}$ & $8.4\times 10^{-11}$\\
        \hline
        Sum of neutrino masses & $\sum m_\nu$ & $0.06$ & $\infty$ & 0.22\\
        \hline
        Dark energy equation of state parameter & $w_0$ & -1 & $\infty$ & 0.062 \\ 
        \hline
        Baryonic feedback strength & $\log T_{\text{AGN}}$ & 7.8 & $\infty$ & $\infty$ \\
        \hline
    \end{tabular}
    \caption{Cosmological parameters varied ($\Lambda \text{CDM} + w_0 + m_\nu$). We use a prior similar to the one used by the \textit{Planck} weak lensing results \cite{Planck2018Lensing} which essentially only restricts $\Omega_bh^2$ and $n_s$ significantly.}
    \label{tab:fiducialpars}
\end{table}

\subsection{Noise Modeling}
\begin{figure}[t]
    \centering
    \includegraphics[width=\textwidth]{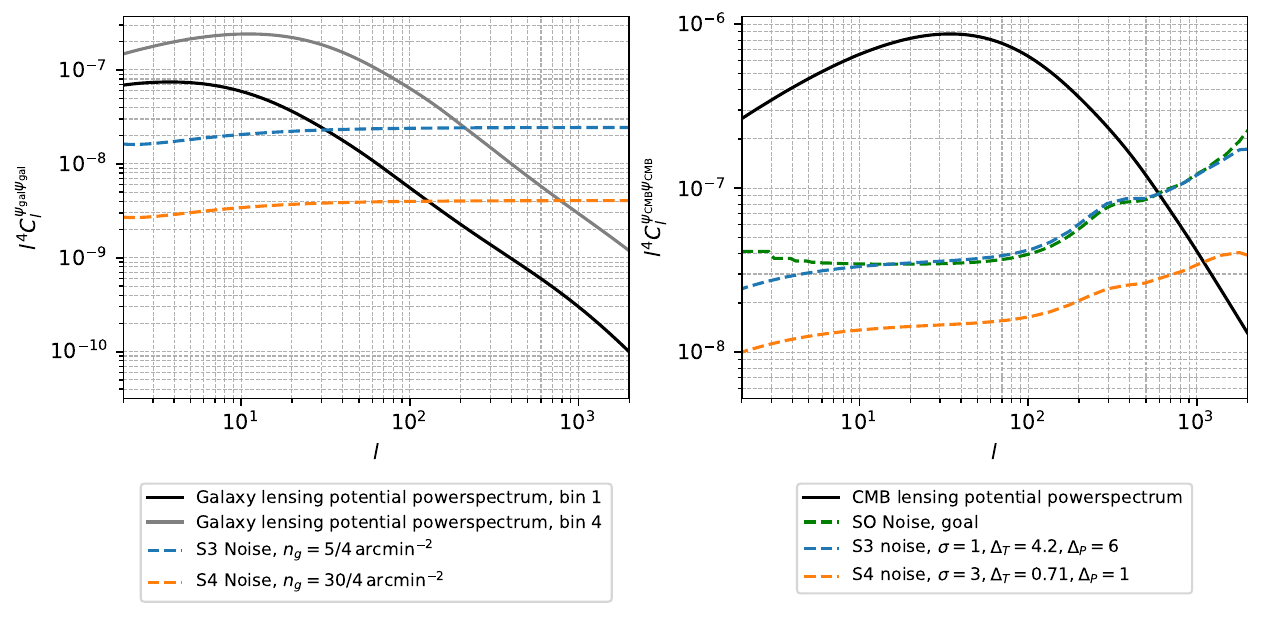}
    \caption{ CMB (right) and galaxy (left) lensing potential power spectra compared to associated experimental noise. Current (stage 3) noise values are displayed as well as near future (some of the stage 4 experiments are already in operation, but expected noise levels are only achieved after several years of integration) (stage 4) noise values. The CMB lensing experiment uses only polarization. CMB noise values are chosen in accordance with \cite{Namikawa_2016}. For comparison we also show the reconstruction noise for the Simons Observatory \cite{Ade2019}. Shear noise values are chosen to be similar to e.g. \textit{Euclid} measurements \cite{laureijs2009} for stage 4 and e.g. KiDS \cite{kuijken2021} for stage 3. Because we have 4 bins the galaxy density is a 4th of the total galaxy density.}
    \label{fig:lpsplusnoise}
\end{figure}

This paper considers noise levels for ``stage 3'' and ``stage 4'' (weak lensing) surveys. Noise properties considered for this analysis can be found in table \ref{tab:noiselevels}, including a list of experiments with comparable noise properties. A comparison of the noise power spectra to the lensing power spectra are shown     in figure \ref{fig:lpsplusnoise}. Effectively, for galaxy lensing, the power spectrum is signal-dominated up until $\ell \sim 200$ (stage 3) and $\ell \sim 700$ (stage 4). For CMB lensing, the power spectrum is signal-dominated for $\ell \sim 300$ (stage 3) and $\ell \sim 1000$ (stage 4). 

\begin{table}[!]
    \centering
    \begin{tabular}{|c|c|c|c|}
    \hline
    \textbf{source} & \textbf{survey stage} & \textbf{noise vals} & \textbf{comparable experiments} \\
    \hline 
    \multirow{2}{*}{CMB} 
        & stage 3 & $\sigma = 1'$, $\Delta_P = 6' \; \mu \text{K}$ & Advanced ACTPol, Simons Observatory \\
    \cline{2-4}
        & stage 4 & $\sigma = 3'$, $\Delta_P = 1' \; \mu \text{K}$ & CMB-S4 \\
    \hline
    \multirow{2}{*}{galaxies} 
        & stage 3 & $\sigma_{\text{rms}} = 0.3$, $n_g = 5 \text{arcmin}^{-2}$ & DES, KiDS \\
    \cline{2-4}
        & stage 4 & $\sigma_{\text{rms}} = 0.3$, $n_g = 30 \text{arcmin}^{-2}$ & LSST, \textit{Euclid} \\
    \hline
    \end{tabular}
    \caption{Noise levels considered for weak lensing of galaxies and the CMB. $\sigma$ (beam width) and $\Delta_P$ (polarization white noise) describe CMB survey specifications, while $\sigma_{\text{rms}}$ (intrinsic galaxy ellipticity) and $n_g$ (observed galaxy density) refer to galaxy shear surveys.}
    \label{tab:noiselevels}
\end{table}

The CMB lensing noise is estimated using a quadratic estimator \cite{Maniyar2021_QE_CMBLensing}.
The parameters characterizing the noise levels are beam width, $\sigma$, polarization white noise, $\Delta_P$, and temperature white noise $\Delta_T$. We take $\Delta_T = \Delta_P / \sqrt 2$ throughout. Galaxy lensing is determined by measuring lensing shear (see appendices \ref{sec:weaklensing} and \ref{sec:shear}). The noise in this measurement is dominated by scale-independent shot noise and has associated noise power spectrum $N_l^{\text{shear}} = \sigma_{\text{rms}}^2 / n_g$, \cite{Bartelmann2001}. $n_g$ is the amount of galaxies observed per unit solid angle. The noise power spectrum for the lensing potential then equals
\begin{equation*}
    N_\ell^{\text{lens. potential}} = \frac{4}{(\ell-1)\ell(\ell+1)(\ell+2)}N_\ell^{\text{shear}}.
\end{equation*}
In all cases we assume that the proportion of the sky surveyed, $f_{\text{sky}}$, equals 0.5.
For parameter constraints, we will consider only stage 4 noise levels.

\subsection{Choice for tomography and redshift uncertainty}
We choose to take 4 bins, each with borders placed so that they contain an equal number of galaxies. To model redshift uncertainty, we convolve each bin distribution with a redshift dependent gaussian uncertainty $\sigma_z = 0.05 (1+z)$. This is representative of the type of uncertainty we expect from surveys such as Euclid \cite{EuclidSciRD}. See figure \ref{fig:galaxyredshiftbinning} for the resulting true distributions of each bin.

\subsection{Other details}
The redshift distribution of the observed galaxies is commonly parameterized as \cite{Bartelmann2001}
\begin{equation}
    n(z) \propto z^a \exp\sbr{-\br{\frac{z}{z_0}}^b}. \label{eq:galzdist}
\end{equation}
We choose the parameter combination $a = 2$, $b = 3/2$ and $z_0 = 0.64$ which is similar to the expected distributions of \textit{Euclid}, (which will probe primarily in the 0.2 - 2.6 redshift range \cite{euclidprep10})
and the LSST mission (which has $a \approx 2$, $b \approx 1$, and $z_0 \approx 0.3$ from predictions for the obtained data \cite{lsstsciencebookchapter3}). For simplicity, we do not implement a tomographic binning of the source galaxies, and instead model the population with a single effective redshift distribution.

Derivatives are calculated with a central difference formula (see appendix \ref{sec:derivatives} for convergence).  To check the accuracy of our results we calculated constraints for similar experimental parameters as in references \cite{Planck2018Lensing, Takada2003, cmbs4sciencebook,neutrinoconstraints} and found good agreement in all cases.

\section{Results}\label{sec:results}
\subsection{Detectability}
\begin{figure}[!]
    \includegraphics[width=\textwidth]{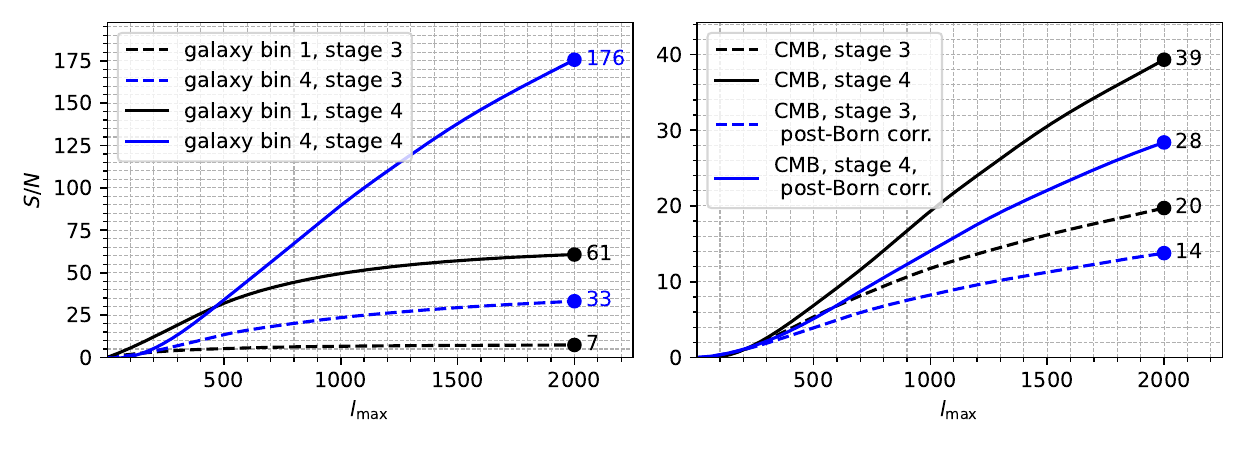}
    \caption{Signal-to-noise ratios for galaxy lensing (left) and CMB lensing (right) bispectra for stage 3 (dashed) and stage 4 (not dashed) surveys as a function of maximum multipole measured. The minimum multipole is always $l_{\text{min}}=2$. The multipole range used for CMB lensing reconstruction is $2\leq \ell \leq 10^4$. Using a more conservative range such as $30\leq \ell \leq 5000$ used to calculate Simons Observatory noise curves don't significantly affect the results.}
    \label{fig:snrplots}
\end{figure}

As expected, lensing power spectra should be detectable at high signal-to-noise. 
The signal-to-noise of the CMB and galaxy lensing bispectra versus the maximum multipole moment measured is shown in figure \ref{fig:snrplots}. In the absence of systematics, shear bispectra can be detected at high signal-to-noise with both stage 3 and stage 4 experiments. We chose to plot the SNR of the bispectra of the individual bins and find that this is also detectable for all bins. CMB lensing bispectra could be  detectable by a stage 3 experiment (see e.g. ref. \cite{Kalaja_2023} for the effect of noise biases) and with high signal-to-noise with a stage 4 experiment. Our results for CMB weak lensing $S/N$ match those of reference \cite{Namikawa_2016}. They also fall in line with e.g. ref. \cite{Schiavone_2024}, where they found detectable skewness of the distance-redshift relation in $\Lambda$CDM due to post-Born effects and non-linear matter formation.

Regarding post-Born corrections, we found that the effect on the galaxy lensing bispectrum $S/N$ is a few percent at most, in agreement with \cite{galpostbornlbscorr}. We thus did not include additional plots for these corrections. For the CMB the $S/N$ is reduced significantly (in agreement with \cite{postborn_pratten_lewis}) but not so severely that they are no longer detectable. Our conclusions thus remain the same: the CMB bispectrum is possibly detectable with stage 3 surveys and definitely detectable with stage 4 surveys in the absence of any further systematics.

\subsection{Parameter Constraints}
\begin{figure}
    \centering
    \includegraphics[width=0.7\textwidth]{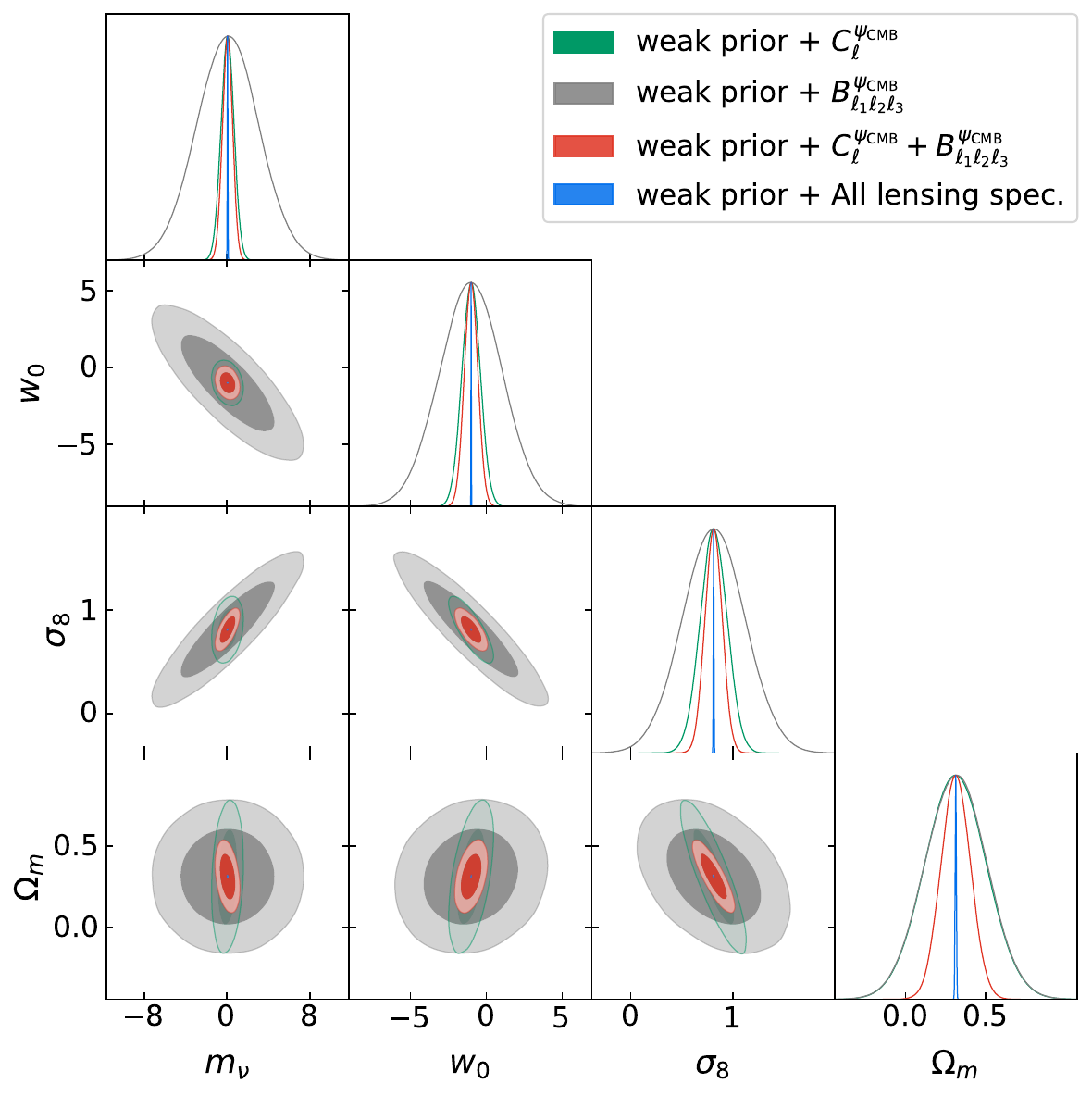}
    \caption{Parameter constraints and confidence ellipses for $\mu_\nu$, $w_0$, $\sigma_8$, and $\Omega_m$. Using ``stage 4'' noise with multipole range from $2$ to $2000$. The colored plots show CMB lensing power and/or bispectrum constraints. We use the weak priors listed in table \ref{tab:fiducialpars}. The confidence ellipses are for $1\sigma$ certainty. They show approximate degeneracies in the information gained from a survey if they are ``stretched''. When the information of ellipses with degeneracies in different directions is combined, the degeneracies are removed and the constraints typically become much better on the relevant parameters. The black plots show the constraints using all lensing information, i.e. CMB lensing spectra, galaxy lensing spectra, and all cross spectra.}
    \label{fig:paramconstraintstightcmb}
\end{figure}

\begin{figure}
    \centering
    \includegraphics[width=0.7\textwidth]{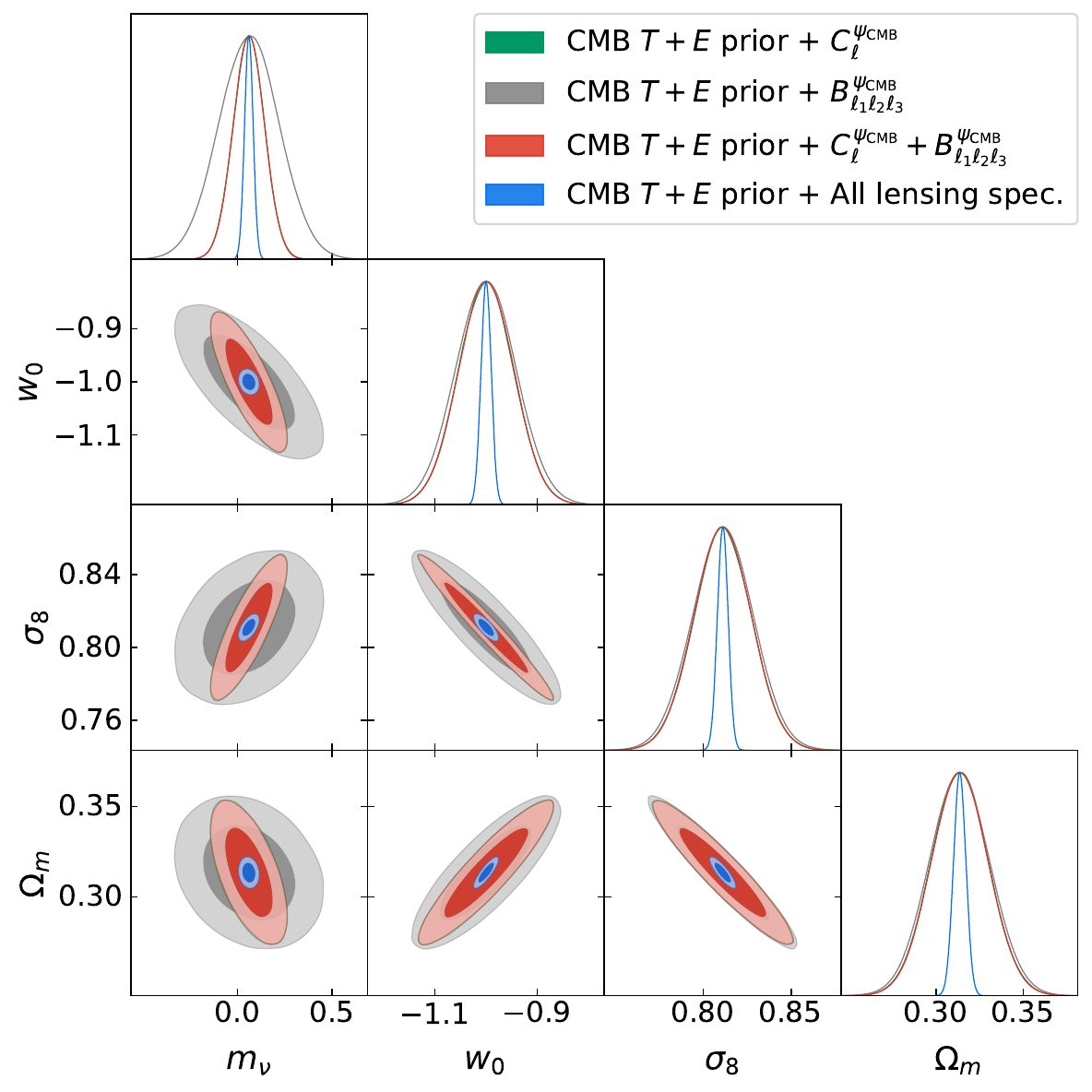}
    \caption{Same as figure \ref{fig:paramconstraintstightcmb}, except we use the CMB temperature and polarization based priors listed in table \ref{tab:fiducialpars}.}
    \label{fig:paramconstraintstightcmbcmbprior}
\end{figure}

\begin{figure}
    \centering
    \includegraphics[width=0.7\textwidth]{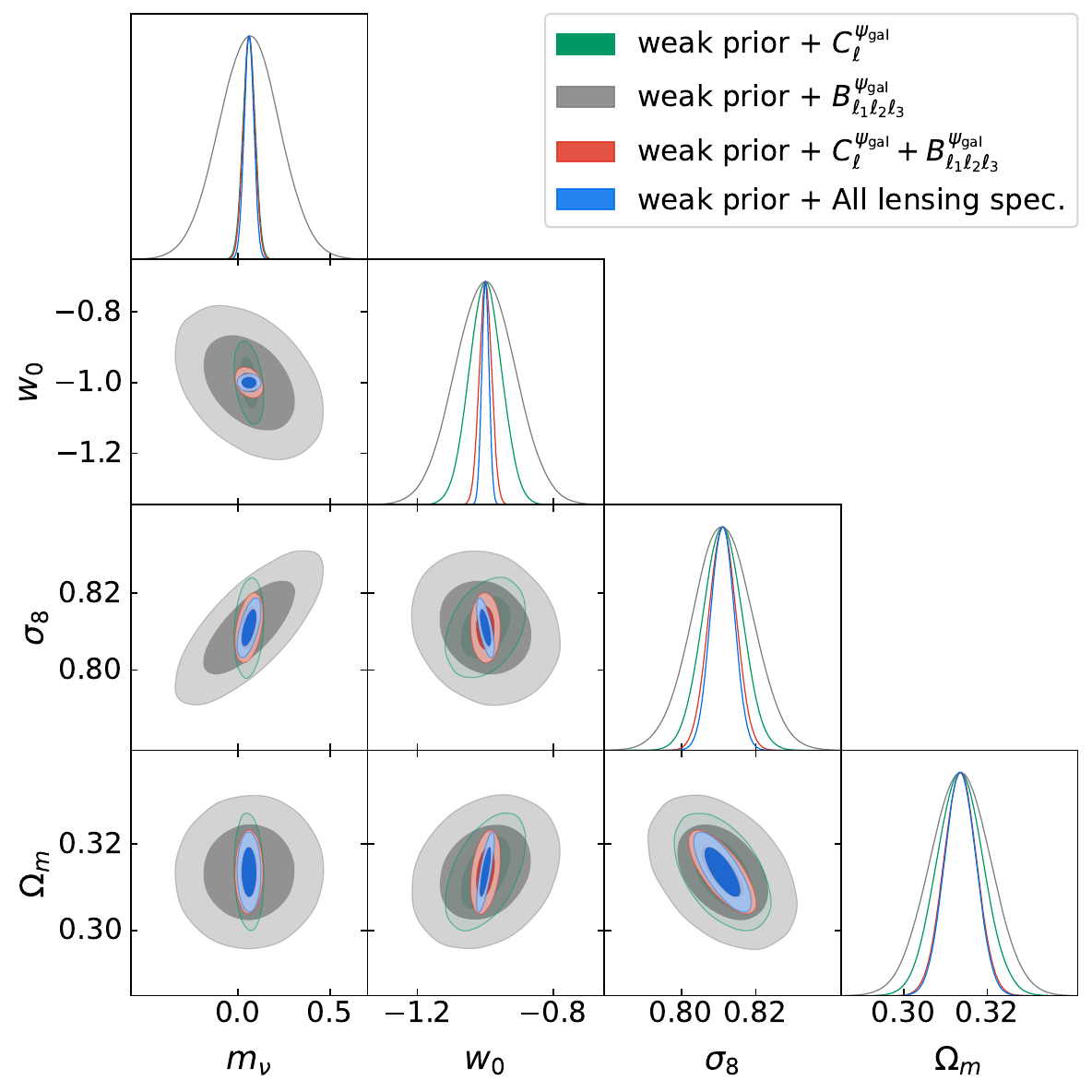}
    \caption{Same as figure \ref{fig:paramconstraintstightcmb}, except here the colored plots show galaxy lensing constraints.}
    \label{fig:paramconstraintstightgal}
\end{figure}

\begin{figure}
    \centering
    \includegraphics[width=0.7\textwidth]{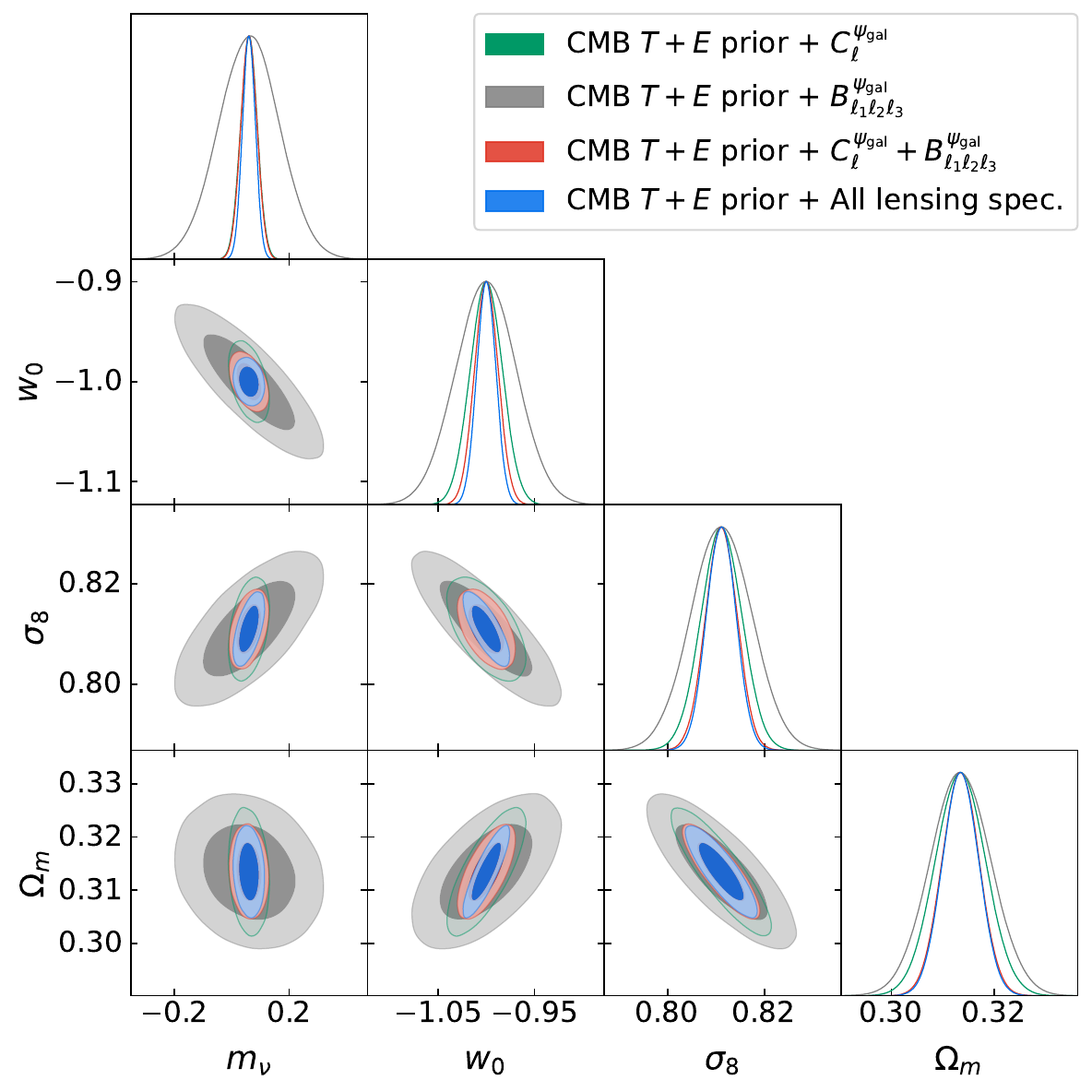}
    \caption{Same as figure \ref{fig:paramconstraintstightcmbcmbprior}, except here the colored plots show galaxy lensing constraints.}
    \label{fig:paramconstraintstightgalcmbprior}
\end{figure}

Next, we consider cosmological parameter constraints. We include both the power spectrum and the bispectrum of the weak lensing. We look at two extensions of $\Lambda CDM$, $m_\nu$ and $w_0$ as well as the derived parameters $\sigma_8$, and $\Omega_m$. Results shown include marginalization over all parameters. Figures \ref{fig:paramconstraintstightcmb} and \ref{fig:paramconstraintstightgal} show confidence ellipses assuming a weak prior from stage 4 CMB and galaxy lensing surveys, respectively. Figures \ref{fig:paramconstraintstightcmbcmbprior} and \ref{fig:paramconstraintstightgalcmbprior} similarly show confidence ellipses in the case of our primary CMB $T+E$ prior. Table \ref{tab:paramconstraintstight} summarizes all forecasted constraints. Further plots and results can be found in appendix \ref{sec:lambdacdmconstraints}.

Constraints are significantly improved by including bispectrum information in nearly all cases when using weak priors. When including the CMB $T+E$ priors, both the CMB lensing power- and bispectra significantly improve prior constraints, however combining them does not lead to significant improvements compared to only using lensing power spectra. For galaxy lensing, the power- and bispectra yield nearly identical improvements compared to the CMB $T+E$ prior. In this case, combining both statistics leads to significantly tighter constraints (in contrast to CMB lensing). When combining CMB and galaxy lensing we see that the bispectrum is slightly less competitive compared to the lensing power spectrum, but can still significantly improve constraints compared to only using the lensing power spectrum.

Furthermore, it is clear that, in general, combining CMB and galaxy lensing surveys also leads to significant improvements in constraining power, even in the case of a strong CMB $T+E$ prior. The improvement is most significant for the neutrino mass constraint.

Besides these obvious findings, we make several other conclusions:
\begin{itemize}
    \item Constraints from all the lensing spectra with weak priors are better than or at least competitive with current constraints from real survey data, see table \ref{tab:currenttoforecastcompare}. Obviously our forecast is optimistic, so it would not be appropriate to say our constraints are a definite improvement. This is also not necessarily expected for constraints coming from essentially only lensing data.
    
    \begin{table}[]
        \centering
        \begin{tabular}{|c|c|c|c|}
            \hline
            Par & Data & Current Constraint & Our Constraint \\
            \hline
            $m_\nu$ & DES Y6 \cite{descollaboration2026darkenergysurveyyear} & $\sum m_\nu < 0.14\,\mathrm{eV}$ (95\% CL) & $0.026\,\mathrm{eV}$ \\
            $w_0$ & DES Y6 \cite{descollaboration2026darkenergysurveyyear}& $\sigma(w_0)\simeq 0.021$ & $0.011$ \\
            $\sigma_8$ & Planck 2018 \cite{planckresults}& $\sigma(\sigma_8)\simeq 0.006$ & $0.0033$ \\
            $\Omega_m$ & DES Y6 \cite{descollaboration2026darkenergysurveyyear}& $\sigma(\Omega_m)\simeq 0.003$ & $0.0041$ \\
            \hline
        \end{tabular}
        \caption{Comparison between some of the best current cosmological constraints and the forecasted constraints obtained from combining all weak lensing data (power spectra, bispectra, galaxy lensing, CMB lensing) with weak external priors.}
        \label{tab:currenttoforecastcompare}
    \end{table}

    \item For lower noise values (in particular for stage 4 noise), the bispectra become increasingly important compared to the power spectra. The lensing bispectra are directly related to the amount of non-Gaussianity in the matter distribution. As noise levels become lower, we are able to measure the matter distribution on small enough scales where nonlinear matter evolution becomes important. Additionally, for galaxy lensing, the window function peaks at later redshifts compared to CMB lensing. The bispectrum is largest at late times, and thus the galaxy lensing bispectrum is significantly easier to detect and is better at constraining parameters than the CMB lensing bispectrum.

    \item The information from bispectra is sensitive to cosmology in a different way than that of the power spectra. The main parameter combinations where approximate degeneracies are broken are $(n_s,\Omega_b h^2),\;(\Omega_b h^2,H_0),\;(\Omega_c h^2,H_0),\;(\Omega_c h^2,\Omega_b h^2)\;(n_s,\Omega_c h^2),\;(m_\nu,H_0),\;(m_\nu,n_s),$ $(A_s,\Omega_b h^2)$ (see also appendix \ref{sec:lambdacdmconstraints}). This means that even if stage 4 surveys are not able to reach the noise levels assumed in this paper, the bispectra will likely still lead to significantly tighter constraints.

    \item Stage 4 galaxy weak lensing surveys generally contain more information than CMB surveys. 
    Comparing the confidence ellipses, it is clear that the two types of surveys depend on the underlying parameters in different ways. This leads to substantially better constraints for $\sigma_8$, $\Omega_m$, $m_\nu$, and $w_0$ when combining CMB and galaxy lensing.

    \item Post-Born corrections do not change the constraints drastically. As expected, they matter less in particular for galaxy lensing and when using the CMB $T+E$ prior. We thus conclude that taking into account post-Born corrections does not lead to significantly different conclusions about the general sensitivity of the lensing spectra.
    
    \item Interestingly, when comparing the post-Born results with the Born-approximate results, although bispectrum-only constraints are generally weaker, their combination with power spectra information instead improves constraints for some parameters. This implies that, at least in some cases, the post-Born contribution to the lensing bispectrum depends on the parameters in a different way than the lensing power spectrum and thus improves the amount of information we can extract from survey data despite making the lensing bispectrum less detectable. 
\end{itemize}

\begin{table}
\centering
\scriptsize
\begin{tabular}{|l|r|rrr|rrr|rrr|}
\hline
\multicolumn{11}{|c|}{weak priors (with and without post-Born corrections)} \\
\hline
&& \multicolumn{3}{c|}{CMB lensing} & \multicolumn{3}{c|}{Gal. lensing} & \multicolumn{3}{c|}{CMB $\times$ Gal. lensing} \\
\hline
Par & prior & $C_\ell$ & $B$ & $C_\ell+B$ & $C_\ell$ & $B$ & $C_\ell+B$ & $C_\ell$ & $B$ & $C_\ell+B$ \\
\hline
$m_\nu$    & 10 & 0.62 & 3.0 & 0.49 (\greenfactor{1.28}) & 0.033 & 0.16 & 0.031 (\greenfactor{1.06}) & 0.028 & 0.15 & 0.026 (\greenfactor{1.05}) \\
\rowcolor{blue!10}
           &    &      & 2.2 & 0.50 (\greenfactor{1.25}) &      & 0.17 & 0.031 (\greenfactor{1.06}) &      & 0.15 & 0.026 (\greenfactor{1.06}) \\
$w_0$      & 10 & 0.60 & 2.1 & 0.44 (\greenfactor{1.36}) & 0.048 & 0.089 & 0.018 (\greenfactor{2.68}) & 0.015 & 0.048 & 0.011 (\greenfactor{1.42}) \\
\rowcolor{blue!10}
           &    &      & 0.97 & 0.37 (\greenfactor{1.64}) &      & 0.088 & 0.018 (\greenfactor{2.70}) &      & 0.051 & 0.011 (\greenfactor{1.41}) \\
$\sigma_8$ & 2.5 & 0.13 & 0.31 & 0.084 (\greenfactor{1.55}) & 0.0053 & 0.0081 & 0.0037 (\greenfactor{1.43}) & 0.0041 & 0.0079 & 0.0032 (\greenfactor{1.29}) \\
\rowcolor{blue!10}
           &     &      & 0.22 & 0.084 (\greenfactor{1.56}) &      & 0.0082 & 0.0037 (\greenfactor{1.43}) &      & 0.0079 & 0.0032 (\greenfactor{1.29}) \\
$\Omega_m$ & 0.67 & 0.19 & 0.19 & 0.092 (\greenfactor{2.07}) & 0.0056 & 0.0073 & 0.0040 (\greenfactor{1.40}) & 0.0052 & 0.0071 & 0.0038 (\greenfactor{1.37}) \\
\rowcolor{blue!10}
           &      &      & 0.19 & 0.085 (\greenfactor{2.26}) &      & 0.0074 & 0.0040 (\greenfactor{1.40}) &      & 0.0068 & 0.0038 (\greenfactor{1.37}) \\
\hline
\multicolumn{11}{|c|}{T+E priors (with and without post-Born corrections)} \\
\hline
&& \multicolumn{3}{c|}{CMB lensing} & \multicolumn{3}{c|}{Gal. lensing} & \multicolumn{3}{c|}{CMB $\times$ Gal. lensing} \\
\hline
Par & prior & $C_\ell$ & $B$ & $C_\ell+B$ & $C_\ell$ & $B$ & $C_\ell+B$ & $C_\ell$ & $B$ & $C_\ell+B$ \\
\hline
$m_\nu$    & 0.22 & 0.083 & 0.16 & 0.082 (\greenfactor{1.01}) & 0.029 & 0.11 & 0.028 (\greenfactor{1.03}) & 0.023 & 0.093 & 0.022 (\greenfactor{1.05}) \\
\rowcolor{blue!10}
           &      &       & 0.16 & 0.082 (\greenfactor{1.01}) &       & 0.11 & 0.028 (\greenfactor{1.03}) &       & 0.095 & 0.022 (\greenfactor{1.05}) \\
$w_0$      & 0.062 & 0.054 & 0.059 & 0.054 (\greenfactor{1.00}) & 0.017 & 0.032 & 0.012 (\greenfactor{1.39}) & 0.013 & 0.028 & 0.0099 (\greenfactor{1.36}) \\
\rowcolor{blue!10}
           &       &       & 0.059 & 0.054 (\greenfactor{1.00}) &       & 0.032 & 0.012 (\greenfactor{1.36}) &       & 0.029 & 0.0099 (\greenfactor{1.36}) \\
$\sigma_8$ & 0.021 & 0.016 & 0.017 & 0.016 (\greenfactor{1.00}) & 0.0042 & 0.0063 & 0.0033 (\greenfactor{1.27}) & 0.0039 & 0.0060 & 0.0030 (\greenfactor{1.30}) \\
\rowcolor{blue!10}
           &       &       & 0.017 & 0.016 (\greenfactor{1.00}) &       & 0.0063 & 0.0033 (\greenfactor{1.27}) &       & 0.0061 & 0.0030 (\greenfactor{1.29}) \\
$\Omega_m$ & 0.018 & 0.016 & 0.017 & 0.016 (\greenfactor{1.00}) & 0.0049 & 0.0059 & 0.0037 (\greenfactor{1.34}) & 0.0048 & 0.0058 & 0.0035 (\greenfactor{1.36}) \\
\rowcolor{blue!10}
           &       &       & 0.017 & 0.016 (\greenfactor{1.00}) &       & 0.0060 & 0.0037 (\greenfactor{1.33}) &       & 0.0058 & 0.0035 (\greenfactor{1.34}) \\
\hline
\end{tabular}
\caption{Parameter constraints for different combinations of weak lensing information. $C$ stands for power spectra, and $B$ for bispectra. All surveys are assumed to be stage 4 surveys with multipole range $2 \leq \ell \leq 2000$. We use the weak and CMB priors described in table \ref{tab:fiducialpars}. The blue colored rows show constraints from bispectra with post-Born corrections. Post-Born corrections are not taken into account for the lensing power spectra as mentioned earlier, so these values are omitted in the blue rows. The values between brackets indicate the fractional improvement in constraining power when combining the power- and bispectra compared to only using the power spectrum. They are color-coded such that greener values indicate stronger improvements.}
\label{tab:paramconstraintstight}
\end{table}


\pagebreak

\section{Discussion and conclusion}\label{sec:discussion}
In this work, we have presented parameter forecasts from CMB lensing and galaxy lensing power and bispectra, considering experimental parameters representative of stage 3 and stage 4 CMB and galaxy surveys. Our analysis shows that while lensing power spectra are detectable at high significance, the inclusion of bispectra generally offers significant improvements when considering parameter constraints when considering a $\Lambda$CDM+ model of the universe. Notably, both CMB and galaxy bispectra are found to be detectable, even with stage 3 experimental noise levels.

The primary impact of incorporating bispectrum information is a tightening of cosmological parameter constraints, particularly for parameters such as $H_0$, $\sigma_8$, and $\Omega_m$, and in breaking key parameter degeneracies. We found that for sufficiently low noise levels, characteristic of stage 4 surveys, the bispectra themselves can offer constraining power comparable to that of the power spectra, due to nonlinear structure formation on small scales being better detected by these future surveys. Furthermore, the combination of CMB and galaxy lensing probes, especially when both power spectra and bispectra are utilized, further enhances constraints and aids in mitigating degeneracies. If all information were to be combined\footnote{And using the CMB $T+E$ prior.} we find particularly tight constraints on neutrino mass and dark energy equation of state in particular, i.e., $\sigma(m_\nu) = 55$ meV and $\sigma(w_0) = 0.017$. However, lensing power spectra alone, even for stage 4 surveys, appear insufficient to competitively constrain an 8-parameter $\Lambda\text{CDM} + w_0 + \sum m_\nu$ model without any priors.

There are a number of limitations to the results presented in this paper. The limitations inherent to the Fisher matrix formalism and to non-physical noise models such as those considered in this paper are already well known. The main limitations specific to this work are:
\begin{itemize}
     \item The fitting formula used for the nonlinear matter bispectrum (from \cite{bispfit}) has an estimated accuracy of only up to about 10 percent. This intrinsic inaccuracy in the model for $B^{\delta}$ will propagate to the lensing bispectra. A correct modeling using e.g. effective field theory would be preferred, and would also allow us to marginalize over bias parameters. 
     
     \item Certain cosmological parameters that affect nonlinear large-scale structure, such as neutrino masses or the dark energy equation of state, likely alter the fitting parameters of the matter bispectrum formula when varied. In the literature, this is currently not considered, which can make the derivatives of the bispectrum with respect to these cosmological parameters less accurate. Matter / lensing bispectrum emulators exist that allow one to vary at least some of the cosmological parameters that we consider, providing a possible solution that should be investigated in future work. An example is the emulator used in \cite{Coulton_2019}, which is based on the Cosmological Massive Neutrino Simulations \cite{Liu_2018} and varies $\sum m_{\nu}$, $A_s$, and $\Omega_m$.
     
     \item We assumed throughout this paper that lensing surveys would use data from multipoles as low as $\ell = 2$.  However, our analytical formulas to calculate lensing spectra utilize the Limber approximation, which is only robust for $\ell \gtrsim 50$. Avoiding the Limber approximation would complicate calculations, requiring knowledge of unequal-time matter power and bispectra\footnote{Technically, this would be computationally much more demanding due to the need to evaluate double integrals instead of single integrals}. Moving beyond the Limber approximation can be achieved using methods such as those explored in reference \cite{Chen:2021vba} and might offer a path to address this. It is also worth mentioning that including foregrounds and other systematics will especially raise the noise for lower multipoles (see e.g. \cite{de_Oliveira_Costa_2006}).

     \item As noted in appendix \ref{sec:fisher}, a Gaussian approximation was used for the covariance of the spectra estimators, i.e. we ignored any contributions not arising from standard Wick contractions. We know that the estimated lensing potential is not Gaussian in general, for two reasons: (i) the matter distribution is not Gaussian. (ii) Even with a Gaussian matter distribution, the quadratic estimator we used will not be Gaussian. Consider, for example, $\langle \hat \phi \hat \phi \hat \phi \rangle$. Each $\hat \phi$ is given as a sum over products of the lensing CMB anisotropies, $T$, $E$, $B$. Writing out each $\hat \phi$ in this way would yield terms such as $\langle TTTTTT\rangle$ (i.e. 6-point functions). These terms are non-zero. In fact, not only would they give products of 2-point functions due to Wick contractions, but also connected $n$-point functions with $n>2$ because the lensed fields are no longer Gaussian even if the original fields are. A next step could thus be to account for these additional contributions, perhaps through N-body and ray-tracing simulations. Notably, such simulations could potentially also allow for a more accurate calculation of the lensing spectra themselves, without relying on fitting formulas for the matter bispectrum or the Limber approximation, depending on simulation accuracy. Obviously, such simulations are expensive. 

     \item Related to the previous point, for a completely correct analysis, we would also need to take into account correlations between a CMB prior and the lensing power and bispectra (see e.g. \cite{schmittfull2018}). Currently, we simply add Fisher matrices; however, taking these correlations into account means computing a covariance matrix for combinations of $T$, $E$, $\phi$ with nonzero entries for correlations between $T$/$E$ and $\phi$. For instance, consider the correlation between the estimator of the temperature power spectrum and the lensing power spectrum at arbitrary multipoles. We will get terms like $\langle TT\hat\phi\hat\phi \rangle$, which will be a sum over 6-point functions like $\langle TT TETE\rangle$ due to our quadratic estimator. These 6-point functions are clearly non-zero, as explained earlier.

     \item Another consideration is that non-Gaussianity of the true lensing field doesn't only change the covariance matrices used to calculate Fisher matrix elements, but also introduces an additional bias to the estimated lensing power spectrum, denoted as $N^{(3/2)}_\ell$ \cite{B_hm_2016, B_hm_2018}. This is because the weights in quadratic estimator from ref. \cite{Hu_2002} are calculated while assuming a Gaussian lensing field. The bias is significant at the percent level for CMB lensing, however it may be larger for cross correlations of the CMB lensing field with low redshift mass tracers (such as galaxy lensing). Fitting $N^{(3/2)}_\ell$ to survey data will loosen constraints for the cosmological parameters.

     \item Other than the lensing power spectrum and bispectrum, there are also other statistics from the lensing field that can be used for parameter constraints. For example, ref. \cite{Liu_2016} showed that the lensing convergence one-point probability distribution function (PDF) and peak counts can tighten constraints in the $\Omega_m$ - $\sigma_8$ plane by $\sim30\%$ over just using the lensing power spectrum for a stage 3 experiment\footnote{Specifically, for the AdvACT experiment \cite{Liu_2016}.}. It is thus clear that future work could combine the power- and bispectrum with these additional statistics to further tighten forecasted constraints.

    \item The CMB lensing noise was estimated via a quadratic estimator. While standard, such estimators may not be strictly optimal, especially with complex foregrounds, non-Gaussianities and on small scales. Estimators that, given our stage 4 noise parameters, perform better do exist; for example, the iterative estimator developed in \cite{Smith_2012} can lead to a significantly better signal-to-noise ratio for the CMB lensing bispectrum, as shown in \cite{Namikawa_2016}. Using such an iterative estimator may also result in the CMB bispectrum being able to significantly improve parameter constraints when combined with lensing power spectra and a CMB $T+E$ prior.

    \item In addition to the general limitations of the Fisher matrix formalism, our implementation also introduces numerical errors due to the use of finite-difference derivatives and interpolation of spectra, leading to inaccuracies at the few percent level (see appendix \ref{sec:derivatives}). These are not expected to qualitatively affect our conclusions.

    \item We did not model foreground contamination such as the cosmic infrared background (CIB), thermal Sunyaev-Zel'dovich (tSZ) effect, or point sources. These foregrounds can bias lensing reconstructions and bispectra, particularly in CMB–galaxy cross-correlations, and should be accounted for in future analyses. Baryonic feedback was also not accounted for. Given that future surveys will probe the matter distribution on smaller scales, this will become increasingly relevant \cite{McCarthy_2022}, particularly for estimating neutrino mass \cite{PhysRevD.101.063534}. Bias due to baryonic feedback can be mitigated, e.g. by using the Principal Component Analysis method \cite{s2024impactcosmologydependencebaryonic}, discarding small scale $T$-modes, or fitting additional parameters that capture the effects of baryonic feedback \cite{McCarthy_2022}. Some of these methods may be able to mitigate baryonic feedback effects without increasing the uncertainty of parameter constraints \cite{McCarthy_2021}. Finally, there are still other well-known sources of uncertainty that we did not take into account, such as photometric redshift errors for galaxies or instrumental systematics.

    \item Although HMCODE2020 models baryonic feedback on the nonlinear matter power spectrum well, it is not guaranteed that we get the same accuracy for modelling feedback effects on the matter bispectrum. This is because it was only tested on the nonlinear matter powerspectrum \cite{Mead_2021}. Clearly, there is good reason to think that an accurate model of the matter powerspectrum will lead to an accurate model of the matter bispectrum due to the fitting equation we use \eqref{eq:matterbispectrum}, but this should be checked in future work. Alternatively, one could use the matter bispectrum emulator from \cite{Burger_2026}, which is specifically made to accurately model baryonic feedback for the scales we expect to probe in a Euclid-like survey.
    
\end{itemize}

Despite these limitations, this work provides valuable insight into the potential of future weak lensing surveys. Addressing these limitations will be crucial for obtaining robust cosmological constraints from upcoming observational data.

\section*{Acknowledgements}
We would like to thank Toshiya Namikawa for providing extra details on his paper on the CMB lensing bispectrum \cite{Namikawa_2016}, which, in particular, allowed us to cross-check (intermediate) results. We would also like to thank Antony Lewis for confirming the correctness of the formulas for post-Born corrections. The authors would like to thank Alex van Engelen and Ian Harrisson for detailed comments and suggestions to improve the manuscript. Finally, the authors thank Luna Jansma, who wrote an MSc thesis on the same subject, which served as the foundation of this study. 

\printbibliography

\appendix

\section{Weak Lensing}\label{sec:weaklensing}
\subsection{Perturbed Photon Paths}
We work in the conformal Newtonian gauge and with natural units. Denoting conformal time and conformal radial distance by $\eta$ and $\chi$, respectively, the perturbed line element in FLRW spacetime is given by
\begin{equation}
    \d s^2 = a^2(\eta)((1 + 2\Psi_N) \d \eta^2 - (1 + 2\Phi_N) \gamma_{ij}\d x^i \d x^j),
\end{equation}
where $\gamma_{ij}$ is the unperturbed line element
\begin{equation}
    \gamma_{ij} = \d x^i \d x^j = \d \chi^2 + f^2_K(\chi)(\d \theta^2 + \sin ^2 \theta \d \phi ^2),
\end{equation}
and $f_K(\chi)$ is the comoving angular diameter distance. We will hereafter only consider a flat universe so that $f_K(\chi) = \chi$. Weak lensing of a point source can be quantified by looking at the deflection field $\mathbf d (\hat{\mathbf n}) = \mathbf \theta_{\text{obs}} - \mathbf \theta_{\text{true}}$, i.e. the (small) difference between the angle at which we see the object and the angle at which we would see the object had no lensing occurred. To first order in $\Psi_N$ and $\Phi_N$, the deflection is given as \cite{dodelson2020modern}
\begin{equation}
    \mathbf d(\hat{\mathbf n}) = -2 \int_0^{\chi_*}\d \chi \frac{\chi_* - \chi}{\chi_*\chi}\nabla_{\hat{\mathbf n}}\Psi(\chi\hat{\mathbf n};\eta_0 - \chi),
\end{equation}
where $\Psi$ is the Weyl Potential, $\Psi := (\Psi_N - \Phi_N)/2$, and $\chi_*$ is the conformal distance to the source. $\nabla_{\hat{\mathbf n}}$ is the derivative along the axes orthogonal to the line of sight. The above equation can be written in terms of the lensing potential, $\psi$, as $\mathbf d (\hat{\mathbf n}) = \nabla_{\hat{\mathbf n}} \psi(\hat{\mathbf n})$, with
\begin{equation}
    \psi(\hat {\mathbf n}) := -2 \int_0^{\chi_*}\d \chi \frac{\chi_* - \chi}{\chi_*\chi}\Psi(\chi\hat{\mathbf n};\eta_0 - \chi).
\end{equation}

If the source is instead distributed over radial distance according to some distribution function $p(\chi)$, with $p(\chi)$ normalized to integrate to 1, the $(\chi_* - \chi)/(\chi_*\chi)$ factor is changed as
\begin{equation*}
    \frac{\chi_*-\chi}{\chi_*\chi} \rightarrow W(\chi) := \int_\chi^{\infty}\d\chi' p(\chi')\frac{\chi'-\chi}{\chi'\chi}.
\end{equation*}
$W(\chi)$ is then called the window function. In the most general case, the lensing potential is thus given by
\begin{equation}
    \psi(\hat {\mathbf n}) := -2 \int_0^{\infty}\d \chi W(\chi)\Psi(\chi\hat{\mathbf n};\eta_0 - \chi).
\end{equation}
The integration limit is sometimes also taken to be the surface of the last scattering, as any window function vanishes after that distance. In the case of CMB lensing we can take $p(\chi')=\delta(\chi'-\chi_*)$, in which case the window function reduces to $H(\chi_* - \chi)(\chi_*-\chi)/(\chi_*\chi)$, with $H(\chi)$ the Heaviside step function.

\subsection{Convergence and Shear}
The magnification matrix is defined as
\begin{equation}
    A_{ij} := \delta_{ij} + \frac{\partial}{\partial n_j}d_i(\hat{\mathbf n}).
\end{equation}
This matrix can be decomposed in the following form, which immediately gives us definitions for the \textbf{convergence}, $\kappa$, \textbf{shear}, $\gamma_1$ and $\gamma_2$, and \textbf{rotation}, $\omega$:
\begin{equation}
    A_{ij}(\hat{\mathbf n}) = \begin{pmatrix}
        1 - \kappa - \gamma_1 & -\gamma_2 + \omega \\
        -\gamma_2 - \omega & 1 - \kappa + \gamma_1
    \end{pmatrix}.
\end{equation}
In the weak lensing regime and under the Born approximation, $A$ is a symmetric matrix by definition, and $\omega$ vanishes; we will ignore it from here on out. Intuitively, $A$ tells you how a small patch in the sky transforms due to lensing. If we change the unlensed direction of a light source by $\delta\hat{\mathbf n}$, then the corresponding change in direction in the lensed image can be calculated as
\begin{equation}
\hat{\mathbf{n}} + \delta \hat{\mathbf{n}} \rightarrow \hat{\mathbf{n}} + \delta \hat{\mathbf{n}} + \mathbf d (\hat{\mathbf{n}} + \delta \hat{\mathbf{n}}) = \hat{\mathbf{n}} + \mathbf d (\hat{\mathbf n}) + \delta \hat{\mathbf n} + A_{ij}(\delta \hat{\mathbf{n}})_j.
\end{equation}
For an image of the sky, $A_{ij}$ thus introduces distortion. Note that $|A_{ij}| = (1 - \kappa)^2 + \omega^2 - |\gamma|^2 = 1 - 2\kappa + O(\kappa^2, \gamma^2, \omega^2)$. We can thus interpret $\kappa$ as telling us about the overall magnification of the source. The $\gamma_i$ represents the area-preserving distortion, i.e. stretching and squeezing in a specific direction.

We can relate $\kappa$ and $\gamma$ directly to the lensing potential as
\begin{equation}
    \kappa = \frac{1}{2}\nabla^2\psi, \quad \gamma_1 = \frac{1}{2}(\partial_{n_1}^2 - \partial_{n_2}^2)\psi, \quad \gamma_2 = \partial_{n_1}\partial_{n_2} \psi.
\end{equation}

It is shown in appendix \ref{sec:shear} that
\begin{equation}
    \gamma := \gamma_1 + i\gamma_2 = \frac{1}{2}\eth_1(\eth_0\psi)
\end{equation}
where the spin raising operator, $\eth_s$ acts on a spin $s$ function defined on $S^2$ to create a spin $s+1$ function.
$\eth_s$ can be written in spherical coordinates $(\theta, \phi)$ as\footnote{We use the physics convention for the definition of $\theta$ and $\phi$ here.}
\begin{equation}
    \eth_s = -\sin^s\theta(\partial_\theta + \frac{i}{\sin\theta})\frac{1}{\sin^s\theta}.
\end{equation}
In this context, a spin $s$ function refers to a function $_sf(\theta, \phi)$ that transforms under any rotation of coordinates by picking up a phase factor $e^{is\alpha}$, with $\alpha$ the angle of the rotation, i.e.
\begin{equation}
    f'(\theta', \phi') = e^{is\alpha}f(\theta, \phi).
\end{equation}
Shear is thus a spin 2 function, which can be checked by noting that rotating a galaxy image stretched and squeezed through weak lensing by $180$ degrees gives the same stretching and squeezing, i.e. the same shear.

The spherical harmonics are eigenfunctions of $\nabla^2$ and the spin raising and lowering operators. Using this property, the corresponding relations in spherical harmonic space can be shown to be 
\begin{equation*}
    \kappa_{lm} = \frac{l(l+1)}{2}\psi_{lm}, \quad \gamma_{lm} = \frac{\sqrt{(l-1)l(l+1)(l+2)}}{2} \psi_{lm}.
\end{equation*}

\section{Weak Lensing Statistics}\label{sec:weaklensstats}
\subsection{Lensing Potential Power spectrum}
The lensing potential can be decomposed into spherical harmonics as
\begin{equation}
    \psi(\hat{\mathbf n}) = \sum_{\ell m} \psi_{\ell m}Y_{\ell m}(\hat{\mathbf n}).
\end{equation}
On the other hand, consider the decomposition of $\Psi$ in Fourier modes with the Fourier convention $\Psi(\mathbf x, \eta)=\int\frac{\d^3 \mathbf k}{(2\pi)^{3}}\Psi(\mathbf k, \eta)e^{i\mathbf k \cdot\mathbf x}$,
\begin{equation}
    \psi(\hat{\mathbf n}) = -2 \int_0^{\chi_*}\d \chi W(\chi) \int\frac{\d^3 \mathbf k}{(2\pi)^{3}}\Psi(\mathbf k, \eta_0 - \chi)e^{i\mathbf k \cdot\hat{\mathbf n}\chi}.
\end{equation}
We can then relate the multipole modes of $\psi$ to the Fourier modes of $\Psi$ through
\begin{gather}
    \psi_{lm} = \langle Y_\ell^m | \psi \rangle = \int\d^2\hat{\mathbf n} Y_\ell^m(\hat{\mathbf n})^* \psi(\hat{\mathbf n})\\ 
    = -2\int\d^2\hat{\mathbf n} Y_l^m(\hat{\mathbf n})^* \int_0^{\chi_*}\d \chi W(\chi) \int\frac{\d^3 \mathbf k}{(2\pi)^{3}}\Psi(\mathbf k, \eta_0 - \chi)e^{i\mathbf k \cdot\hat{\mathbf n}\chi}
\end{gather}
Now, define the power spectrum as
\begin{equation}
    \langle \Psi(\mathbf k, \eta)\Psi(\mathbf k',\eta')\rangle = \frac{2\pi^2}{k^3}P_\Psi(k, \eta, \eta')\delta(\mathbf k + \mathbf k'),
\end{equation}
with $\eta$ denoting the conformal time. This gives 
\begin{equation}
    \langle \psi(\hat{\mathbf n})\psi(\hat{\mathbf n}') \rangle = 4\int_0^{\chi_*}\d \chi\int_0^{\chi_*}\d \chi' W(\chi)W(\chi')\int \frac{\d^3\mathbf k}{(2\pi)^6}\frac{2\pi^2}{k^3}P_\psi(k,\eta_0 - \chi, \eta_0-\chi')e^{i\mathbf k\cdot\hat {\mathbf n}\chi}e^{-i\mathbf k\cdot\hat {\mathbf n}'\chi'},
\end{equation}
where we used that $\eta = \eta_0 - \chi$ along the unperturbed photon path (this is known as the Born approximation), with $\eta_0$ the time at which the light ray hits Earth. 
We can use the result
\begin{equation}
    e^{i\mathbf k \cdot \hat {\mathbf n}\chi}=4\pi\sum_{\ell m}i^\ell j_\ell(k\chi )Y_\ell^m(\hat{\mathbf n})^*Y_\ell^m(\hat{\mathbf k}) = 4\pi\sum_{\ell m}i^\ell j_\ell(k\chi )Y_\ell^m(\hat{\mathbf n})Y_\ell^m(\hat{\mathbf k})^*, \label{eq:complexexp}
\end{equation}
where $j_l$ is the spherical Bessel function, to rewrite the above equation. Using both versions of the identity above, we immediately get a factor $Y_\ell^m(\hat{\mathbf k})Y_{\ell'}^{m'}(\hat{\mathbf k})^*$ in our integral. We can factor the differential element of $\d^3\mathbf k$ into a radial and angular part as $k^2\d k\d^2\Omega_k$, with $\Omega_k$ the solid angle, to apply the orthonormality condition of the spherical harmonics. Note that we take the same sequence of steps a number of times in other parts of the derivations of the lensing spectra. We thus obtain
\begin{gather}
    \langle \psi(\hat{\mathbf n})\psi(\hat{\mathbf n}') \rangle = 4(4\pi)^2\sum_{ll'mm'}i^{l-l'}\int_0^{\chi_*}\d \chi\int_0^{\chi_*}\d \chi' W(\chi)W(\chi')\\\times\int \frac{k^2\d k}{(2\pi)^6}\frac{2\pi^2}{k^3}j_\ell(k\chi)j_{\ell'}(k\chi')P_\psi(k,\eta_0 - \chi, \eta_0-\chi')Y_{\ell m}(\hat{\mathbf n})Y_{\ell'm'}(\hat{\mathbf n}')^*\delta_{\ell \ell'}\delta_{mm'}. \label{eq:lenspotcorrexpanded}
\end{gather}
The angular power spectrum is defined similarly to the power spectrum, i.e.
\begin{equation}
    \langle \psi_{\ell m}\psi_{\ell'm'}^* \rangle = \delta_{\ell \ell'}\delta_{mm'}C_\ell^\psi.
\end{equation}
Note that the correlation is independent of $m$ and $m'$.
We can thus read from equation \ref{eq:lenspotcorrexpanded} that
\begin{gather}
    C_l^\psi = 4(4\pi)^2\int_0^{\chi_*}\d \chi\int_0^{\chi_*}\d \chi' W(\chi)W(\chi')\int \frac{k^2\d k}{(2\pi)^6}\frac{2\pi^2}{k^3}j_\ell(k\chi)j_{\ell}(k\chi')P_\psi(k,\eta_0 - \chi, \eta_0-\chi'),
\end{gather}
which can be simplified to
\begin{gather}
    C_\ell^\psi = \frac{2}{\pi^2}\int_0^{\chi_*}\d \chi\int_0^{\chi_*}\d \chi' W(\chi)W(\chi')\int k^2\d kj_\ell(k\chi)j_{\ell}(k\chi')\frac{P_\psi(k,\eta_0 - \chi, \eta_0-\chi')}{k^3}.
\end{gather}

To further evaluate the integral we will make the Limber approximation. The Bessel functions peak sharply at $l=x$\footnote{Some sources use $x\approx l+1/2$ instead, which is slightly more accurate for larger scales (low $l$) and slightly less accurate for smaller scales.}, with the peak being increasingly sharp for higher $l$. Similarly to $\delta(x-x_0)f(x)=\delta(x-x_0)f(x_0)$, we thus take $j_l(k\chi)f(k)\approx j_l(k\chi)f(l/\chi)$. The Bessel functions satisfy an orthogonality condition,
\begin{equation}
    \int k^2\d k j_\ell(k\chi)j_\ell(k\chi') = \frac{\pi}{2\chi^2}\delta(\chi-\chi').
\end{equation}
In combination with the Limber approximation, we thus find
\begin{equation}
    \int k^2\d k j_\ell(k\chi)j_\ell(k\chi')f(k) \approx \frac{\pi}{2\chi^2}\delta(\chi-\chi')f(\ell/\chi).
\end{equation}
It allows us to write the Limber-approximate angular spectrum as
\begin{gather}
    C^\psi_{\ell} = \frac{2}{\pi^2}\int_0^{\chi_*}\d \chi\int_0^{\chi_*}\d \chi' W(\chi)W(\chi') \frac{\pi}{2\chi^2}\delta(\chi-\chi')\frac{\chi^3}{\ell^3}P_\psi(\ell/\chi,\eta_0 - \chi, \eta_0-\chi')\\
    = \frac{1}{\ell^3\pi}\int_0^{\chi_*}\chi\d \chi W(\chi)^2 P_\psi(\ell/\chi,\eta_0 - \chi, \eta_0-\chi). \label{eq:lpsintermsofpsi}
\end{gather}

\subsection{Lensing potential bispectrum}
The derivation of the bispectrum proceeds similarly to that of the power spectrum. We aim to compute the bispectrum of the lensing potential fields of 3 (possibly distinct sources), $\psi_1$, $\psi_2$, $\psi_3$. 
\begin{gather*}
    \langle (\psi_1)_{\ell_1m_1}(\psi_2)_{\ell_2m_2}(\psi_3)_{\ell_3m_3} \rangle = 
    \prod_i\br{-2\int\d^2\hat{\mathbf n_i}(Y^{m_i}_{\ell_i}(\hat{\mathbf n_i}))^*\int_0^{\chi_*}\d\chi_i W_i(\chi_i)\int\frac{\d^3\mathbf k_i}{(2\pi)^{3}}e^{i\mathbf k_i \cdot\hat{\mathbf n_i}\chi_i}} \\
    \times \langle \prod_i\Psi(\mathbf k_i, \eta_0 - \chi_i) \rangle.
\end{gather*}
Defining the bispectrum of the gravitational potential as
\begin{equation*}
    \langle \prod_{i=1,2,3}\Psi(\mathbf k_i, \eta_0 - \chi_i) \rangle = (2\pi)^3\delta(\mathbf k_1 + \mathbf k_2 + \mathbf k_3)B^\Psi(\{k_i\}, \{\eta_0-\chi_i\}),
\end{equation*}
we rewrite the lensing potential bispectrum as
\begin{gather*}
    \langle (\psi_1)_{\ell_1m_1}(\psi_2)_{\ell_2m_2}(\psi_3)_{\ell_3m_3} \rangle = 
    \prod_i\br{-2\int\d^2\hat{\mathbf n_i}(Y^{m_i}_{\ell_i}(\hat{\mathbf n_i}))^*\int_0^{\chi_*}\d\chi_i W_i(\chi_i)\int\frac{\d^3\mathbf k_i}{(2\pi)^{3}}e^{i\mathbf k_i \cdot\hat{\mathbf n_i}\chi_i}}\\ \times (2\pi)^3\delta(\mathbf k_1 + \mathbf k_2 + \mathbf k_3)B^\Psi(\{k_i\}, \{\eta_0-\chi_i\}).
\end{gather*}
Now using equation \eqref{eq:complexexp} to rewrite the complex exponential:
\begin{gather*}
    \langle (\psi_1)_{\ell_1m_1}(\psi_2)_{\ell_2m_2}(\psi_3)_{\ell_3m_3} \rangle \\ = 
    \prod_i\br{-2\int\d^2\hat{\mathbf n_i}(Y^{m_i}_{\ell_i}(\hat{\mathbf n_i}))^*\int_0^{\chi_*}\d\chi_i W_i(\chi_i)\int\frac{\d^3\mathbf k_i}{(2\pi)^{3}}4\pi\sum_{\ell m}i^{\ell}j_{l}(k_i\chi_i)Y_{\ell}^{m}(\hat{\mathbf n_i})Y_{\ell}^{m}(\hat{\mathbf k_i})^*}\\
    \times (2\pi)^3\delta(\mathbf k_1 + \mathbf k_2 + \mathbf k_3)B^\Psi(\{k_i\}, \{\eta_0-\chi_i\}) \\
    = \sbr{\prod_i\br{-2\int_0^{\chi_*}\d\chi_i W_i(\chi_i)\int\frac{\d^3\mathbf k_i}{(2\pi)^{3}}4\pi i^{\ell_i}j_{\ell_i}(k_i\chi_i)Y_{\ell_i}^{m_i}(\hat{\mathbf k_i})^*}} (2\pi)^3\delta(\mathbf k_1 + \mathbf k_2 + \mathbf k_3)B^\Psi(\{k_i\}, \{\eta_0-\chi_i\}).
\end{gather*}
We can write the 3D Dirac delta function in terms of spherical harmonics as
\begin{equation*}
    \delta(\mathbf k_1 + \mathbf k_2 + \mathbf k_3) = 8\int\d^3\mathbf x\prod_{i=1,2,3}\br{ \sum_{\ell_jm_j} i^{\ell_j}j_{\ell_j}(k_ix)Y_{\ell_j}^{m_j}(\hat{\mathbf k_i}) Y_{\ell_j}^{m_j}(\hat{\mathbf x})^* }.
\end{equation*}
This results in
\begin{gather*}
    \langle (\psi_1)_{\ell_1m_1}(\psi_2)_{\ell_2m_2}(\psi_3)_{\ell_3m_3} \rangle= \prod_i\br{-2\int_0^{\chi_*}\d\chi_i W_i(\chi_i)\int\frac{\d^3\mathbf k_i}{(2\pi)^{3}}4\pi i^{\ell_i}j_{\ell_i}(k_i\chi_i)Y_{\ell_i}^{m_i}(\hat{\mathbf k_i})^*} \\
    \times (2\pi)^38\int\d^3\mathbf x\prod_{i}\br{ \sum_{\ell m} i^{\ell}j_{\ell}(k_ix)Y_{\ell}^{m}(\hat{\mathbf k_i}) Y_{\ell}^{m}(\hat{\mathbf x})^* }B^\Psi(\{k_i\}, \{\eta_0-\chi_i\})\\
    = (2\pi)^38\int\d^3\mathbf x\prod_i\br{-2\int_0^{\chi_*}\d\chi_i W_i(\chi_i)\int\frac{k_i^2\d k_i}{(2\pi)^{3}}4\pi (-1)^{\ell_i}j_{\ell_i}(k_i\chi_i)j_{\ell_i}(k_ix) Y_{\ell_i}^{m_i}(\hat{\mathbf x})^*} B^\Psi(\{k_i\}, \{\eta_0-\chi_i\})
\end{gather*}
The angular part of the $\mathbf x$ integral can be evaluated using the identity
\begin{gather*}
    \int \mathrm{d}\Omega_{\hat{n}} \, Y_{\ell_1 m_1}^*(\hat{x}) Y_{\ell_2 m_2}^*(\hat{n}) Y_{\ell_3 m_3}^*(\hat{n}) = (-1)^{m_1 + m_2 + m_3} 
    \int \mathrm{d}\Omega_{\hat{n}} \, Y_{\ell_1 -m_1}(\hat{n}) Y_{\ell_2 -m_2}(\hat{n})Y_{\ell_3 -m_3}(\hat{n}) \\
    = (-1)^{m_1 + m_2 + m_3} 
    \sqrt{\frac{(2\ell_1 + 1)(2\ell_2 + 1)(2\ell_3 + 1)}{4\pi}} \begin{pmatrix} \ell_1 & \ell_2 & \ell_3 \\ 0 & 0 & 0 \end{pmatrix}
    \begin{pmatrix} \ell_1 & \ell_2 & \ell_3 \\ -m_1 & -m_2 & -m_3 \end{pmatrix} \equiv A_{\mathbf{l}}^{\mathbf{m}},
\end{gather*}
giving
\begin{gather*}
    \langle (\psi_1)_{\ell_1m_1}(\psi_2)_{\ell_2m_2}(\psi_3)_{\ell_3m_3} \rangle 
    = (2\pi)^3 8 A_{\mathbf l}^{\mathbf m}\int x^2\d x\prod_i\br{-2\int_0^{\chi_*}\d\chi_i W_i(\chi_i)\int\frac{k_i^2\d k_i}{(2\pi)^{3}}4\pi (-1)^{\ell_i}j_{\ell_i}(k_i\chi_i)j_{\ell_i}(k_ix)} \\ \times B^\Psi(\{k_i\}, \{\eta_0-\chi_i\}).
\end{gather*}
Now applying the Limber approximation again:
\begin{gather*}
    \langle (\psi_1)_{\ell_1m_1}(\psi_2)_{\ell_2m_2}(\psi_3)_{\ell_3m_3} \rangle 
    = (2\pi)^3 8 A_{\mathbf l}^{\mathbf m}\int x^2\d x\prod_i\br{-2\int_0^{\chi_*}\d\chi_i W_i(\chi_i)\frac{1}{(2\pi)^{3}}\frac{\pi}{2\chi_i^2}\delta(x-\chi_i)4\pi (-1)^{\ell_i}} \\ \times B^\Psi(\{\ell_i/\chi_i\}, \{\eta_0-\chi_i\}) \\
    = (2\pi)^3 8 A_{\mathbf l}^{\mathbf m}\int \chi^2\d \chi\prod_i\br{-2 W_i(\chi)\frac{1}{(2\pi)^{3}}\frac{\pi}{2\chi^2}4\pi (-1)^{\ell_i}} B^\Psi(\{\ell_i/\chi\}, \eta_0-\chi).
\end{gather*}
Finally, we aim to rewrite the above in terms of the angular bispectrum of the lensing potential. 

The definition for the bispectrum of any set of randomly distributed spherical harmonic components $X^k_{\ell m}$ is \cite{Hu2000}
\begin{equation*}
    \langle (X_1)_{\ell_1m_1}(X_2)_{\ell_2m_2}(X_3)_{x_3m_3} \rangle = \begin{pmatrix}
        \ell_1 & \ell_2 & \ell_3 \\ m_1 & m_2 & m_3
    \end{pmatrix}
    B_{\ell_1\ell_2\ell_3}^{X_1X_2X_3}.
\end{equation*}
Note the independence on $m_i$, this necessarily follows from statistical isotropy. 
If $m_1+m_2+m_3\neq 0$, the associated Wigner-3j symbol vanishes and the bispectrum is set to zero.
Also note that in this definition we have immediately generalized to include cross correlation between different fields $X_1$, $X_2$, $X_3$. This is relevant when we look at cross-correlations between CMB and galaxy lensing.
Using the above definition and the symmetry property
\begin{equation*}
    \begin{pmatrix}
        \ell_1 & \ell_2 & \ell_3\\
        -m_1 & -m_2 & -m_3
      \end{pmatrix}
      =
      (-1)^{\ell_1+\ell_2+\ell_3}
      \begin{pmatrix}
        \ell_1 & \ell_2 & \ell_3\\
        m_1 & m_2 & m_3
      \end{pmatrix},
\end{equation*}
we find
\begin{gather*}
    B^{\psi_1\psi_2\psi_3}_{\ell_1\ell_2\ell_3}
    = (-1)^{\ell_1+\ell_2+\ell_3}
    \sqrt{\frac{(2\ell_1 + 1)(2\ell_2 + 1)(2\ell_3 + 1)}{4\pi}} \begin{pmatrix} \ell_1 & \ell_2 & \ell_3 \\ 0 & 0 & 0 \end{pmatrix}
    (2\pi)^3 8 \\
    \times \int \chi^2\d \chi\prod_i\br{-2 W_i(\chi, \chi_*)\frac{1}{(2\pi)^{3}}\frac{\pi}{2\chi^2}4\pi (-1)^{\ell_i}} B^\Psi(\{\ell_i/\chi\}, \{\eta_0-\chi\}),
\end{gather*}
where we were able to drop the $(-1)^{m_1+m_2+m_3}$ factor due to the bispectrum vanishing if that sum does not equal $0$, as mentioned earlier. When all $m_i$ equal zero, the Wigner 3j-symbol gains a number of useful properties.
In particular, it vanishes if $\ell_1+\ell_2+\ell_3$ is odd, meaning we can drop the $(-1)^{\ell_1+\ell_2+\ell_3}$ factor. Additionally, cancelling common factors then gives
\begin{align}
    B^{\psi_1\psi_2\psi_3}_{\ell_1\ell_2\ell_3}
    =
    -\sqrt{\frac{(2\ell_1 + 1)(2\ell_2 + 1)(2\ell_3 + 1)}{4\pi}} \begin{pmatrix} \ell_1 & \ell_2 & \ell_3 \\ 0 & 0 & 0 \end{pmatrix} 8 \int \frac{\d \chi }{\chi^4} W_1(\chi)W_2(\chi)W_3(\chi) B^\Psi(\{l_i/\chi\}, \eta_0-\chi). \label{eq:lbsintermsofpsi}
\end{align}

\subsection{Gravitational potential spectra in terms of matter spectra}
We can rewrite equations \ref{eq:lpsintermsofpsi} and \ref{eq:lbsintermsofpsi} in terms of the matter spectra instead of the $\psi$ spectra using the Poisson equation. This allows us to numerically evaluate these lensing spectra using CAMB. The density contrast is defined as
\begin{equation}
    \delta(\mathbf x) := \frac{\rho(\mathbf x)-\bar \rho}{\bar\rho},
\end{equation}
and the matter spectra are defined in terms of the Fourier transformed density contrast $\delta(\mathbf k)$ as 
\begin{gather*}
    \langle \delta(\mathbf k, \eta) \delta(\mathbf k', \eta)^* \rangle = (2\pi)^3\delta(\mathbf k - \mathbf k')P^\delta(\mathbf k, \eta),\\
    \langle \delta(\mathbf k_1, \eta) \delta(\mathbf k_2, \eta) \delta(\mathbf k_3, \eta) \rangle = (2\pi)^3\delta(\mathbf k_1 + \mathbf k_2 + \mathbf k_3)B^\delta(k_1, k_2, k_3, \eta).
\end{gather*}
The mean matter density of the universe, \( \bar{\rho} \) is given as 
$$
\bar{\rho}(\eta) = \frac{3 \Omega_m H_0^2}{8 \pi G}\frac{1}{a(\eta)^3},
$$
where $a(\eta)$ is the only time-dependent factor on the right-hand side.
The Poisson equation relates $\Psi$ to the density contrast as \cite{dodelson2020modern}
\begin{equation}
    \nabla^2\Psi(\mathbf x) = 4\pi G a^2\br{\frac{3\Omega_m H_0^2}{8\pi G}\frac{1}{a^3}}\delta(\mathbf x) = \frac{3\Omega_m H_0^2}{2} \frac{1}{a}\delta(\mathbf x) \implies \Psi(k, \eta) = -\frac{3\Omega_m H_0^2}{2}\frac{1}{a}\frac{\delta(k, \eta)}{k^2},
\end{equation}
where $\Psi(k,\eta)$ and $\delta(k,\eta)$ are functions in Fourier space. For the power- and bispectra, we find
\begin{gather*}
    \langle \Psi(\mathbf k, \eta)\Psi^*(\mathbf k',\eta)\rangle = \frac{2\pi^2}{k^3}C^\Psi(k, \eta)\delta(\mathbf k - \mathbf k') \implies C^\Psi(k,\eta) = \frac{1}{k}(9\Omega_m^2H_0^4\pi) \frac{1}{a^2}C^\delta(k,\eta),\\
    \langle \Psi(\mathbf k_1, \eta)\Psi(\mathbf k_2,\eta)\Psi(\mathbf k_3,\eta)\rangle = -(2\pi)^3\delta(\mathbf k_1 + \mathbf k_2 + \mathbf k_3)B^\Psi(k_1,k_2,k_3, \eta)\\ 
    \implies B^\Psi(k_1,k_2,k_3, \eta) = -\frac{1}{k_1^2k_2^2k_3^2}\br{\frac{3\Omega_m H_0^2}{2}}^3 \frac{1}{a^3}B^\delta(\{k_i\}, \eta).
\end{gather*}
Finally, we obtain:
\begin{empheq}[box=\fbox]{align*}
    C^{\psi_X\psi_Y}_\ell
    & = \frac{9}{\ell^4}\Omega_m^2H_0^4\int_0^{\chi_*} \chi^2 \d\chi a(\eta_0-\chi)^{-2}W_X(\chi)W_Y(\chi)P^\delta(\ell/\chi,\eta_0-\chi),\\
    B^{\psi_X\psi_Y\psi_Z}_{\ell_1\ell_2\ell_3} &= \sqrt{\frac{(2\ell_1 + 1)(2\ell_2 + 1)(2\ell_3 + 1)}{4\pi}} \begin{pmatrix} \ell_1 & \ell_2 & \ell_3 \\ 0 & 0 & 0 \end{pmatrix} \frac{27}{\ell_1^2\ell_2^2\ell_3^2}\Omega_m^3H_0^6\\
    & \quad \times \int \chi^2\d \chi a(\eta_0-\chi)^{-3}W_X(\chi)W_Y(\chi)W_Z(\chi)  B^\delta(\{\ell_i/\chi\}, \eta_0-\chi).
\end{empheq}

\section{Fisher Matrix Analysis} \label{sec:fisher}
\subsection{Determining uncertainty in experimental parameters}

The Fisher matrix formalism is used to find a lower bound on the constraints we can place on experimental parameters. It combines the Cramer-Rao Inequality \cite{casella2002statistical} with the assumption that we have unbiased estimators following a Gaussian distribution \cite{dodelson2020modern}. In particular, denoting the Fisher matrix by $F_{\theta_i\theta_j}$, the parameters as $\theta_i$, and their estimators as $\hat\theta_i$, it can be shown that
\begin{equation}
    \text{Var}(\hat\theta_i) \geq \br{F^{-1}}_{\theta_i\theta_i}.
\end{equation}
In the case of $n$ measurements whose outcomes are realizations of random variables $x_i$, each with associated mean $\mu_{x_i}$, the Fisher matrix is given as
\begin{equation}
    F_{\theta_i\theta_j} := \sum_{p, q=1}^n \frac{\partial\mu_{x_p}}{\partial\theta_i}(\tilde\theta_k)(\text{Cov}^{-1})_{x_px_q}(\tilde \theta_k)\frac{\partial\mu_{x_q}}{\partial\theta_j}(\tilde\theta_k),
\end{equation}
where Cov is the covariance matrix associated with the random vector $(x_1, ..., x_n)$, $\text{Cov}_{x_px_q}:=\text{Cov}(x_p,x_q)$. The derivative of the mean measurement outcomes and measurement covariances will in general, depend on the true, unknown values of the experimental parameters. We therefore evaluate these quantities for our best guess of the experimental parameters given some outside information. These are known as the ``fiducial'' values.


\subsection{Fisher matrices for power- and bispectra with multiple tracers}
For power spectra, the definition of the Fisher matrix gives 
\begin{equation*}
    F_{\alpha\beta} = \sum_{l_{min}\leq \ell, \ell' \leq \ell_{max}}\sum_{[XY][X'Y']} \partial_\alpha C_{\ell}^{XY} \br{\Cov^{-1}}^{XY, X'Y'}_{\ell, \ell'} \partial_\beta C^{X'Y'}_{\ell'}.
\end{equation*}
Here, the covariance matrix is given as
\begin{equation*}
    \Cov^{XY, X'Y'}_{\ell, \ell'} = \langle \hat C^{XY}_{\ell} \hat C^{X'Y'}_{\ell'} \rangle.
\end{equation*}
The estimator of the power- or bi-spectrum is the product of the estimators of the appropriate tracers, e.g. $\hat C^{XY}_{\ell} = \hat X(\ell) \hat Y(\ell)$. The sum over $[XY]$ and $[X'Y']$ denotes a sum over possible tracer combinations \textit{without} counting permutations of tracer configuration. This is because permutations are not distinct signals, i.e. $\hat X(\ell) \hat Y(\ell) = \hat Y(\ell) \hat X(\ell)$. In fact, if we were to count these permutations as distinct signals, we would get identical columns and/or rows in our covariance matrix, making inversion impossible:
\begin{equation*}
    \langle \hat X(\ell) \hat Y (\ell) \hat X'(\ell') \hat Y' (\ell') \rangle = \langle \hat Y (\ell) \hat X (\ell) \hat X'(\ell') \hat Y' (\ell') \rangle, \quad \forall X', Y', \ell'. 
\end{equation*}

To evaluate the covariance matrix, we again assume that the estimators are Gaussian distributed so that we can apply a Wick contraction, as is commonly done \cite{tegmark1997cosmic}. In this case we get
\begin{equation*}
    \Cov^{XY, X'Y'}_{\ell, \ell'} = \frac{1}{2\ell+1}\delta_{\ell \ell'}\br{\tilde C^{XX'}_\ell \tilde C^{YY'}_\ell + \tilde C^{XY'}_\ell \tilde C^{YX'}_\ell}.
\end{equation*}
Two remarks are in order.
\begin{enumerate}
    \item By definition
    \begin{equation*}
        \langle X_{\ell m}X'_{\ell'm'} \rangle = (2\pi)^3 \delta_{\ell\ell'}\delta_{mm'}C_\ell^{XX'},
    \end{equation*}
    so you can average over the measurements done for different values of $m$, i.e. $X_{\ell(-\ell)}$, $X_{\ell(-\ell+1)}$, $\cdots$, $X_{\ell(\ell-1)}$, $X_{\ell \ell}$. This results in the $(2\ell+1)^{-1}$ factor in the power spectrum covariance. 
    \item The tilde is used to denote the power spectrum as calculated earlier, plus the noise power spectrum, $N^{XX'}_\ell$, which is the power spectrum of the noise associated with the estimator of the field. This is where experimental noise is incorporated into the calculation. 
\end{enumerate}
Under the Gaussian approximation, the covariance matrix vanishes except for $3\times 3$ block matrices (in the case of 2 tracers) on the diagonal. The Fisher matrix is then
\begin{equation*}
    F_{\alpha\beta} = \sum_{\ell}\sum_{[XY][X'Y']} \partial_\alpha C_{\ell}^{XY} \br{\Cov^{-1}_\ell}^{XY, X'Y'} \partial_\beta C^{X'Y'}_{\ell}.
\end{equation*} 
$\Cov^{-1}_\ell$ here denotes the inverse of the block matrix at $l$.

Next, we consider the Fisher matrix for bispectra measurements.
\begin{equation*}
    F_{\alpha\beta} = \sum_{\text{distinct signals}}\sum_{\text{distinct signals prime}} B^{XYZ}_{\ell_1\ell_2\ell_3}\br{\Cov^{-1}}^{XYZ,X'Z'Y'}_{\ell_1\ell_2\ell_3,\ell_1'\ell_2'\ell_3'}B^{X'Y'Z'}_{\ell_1'\ell_2'\ell_3'}.
\end{equation*} 
Counting only distinct signals requires more care compared to the power spectra. The rule is that $B^{XYZ}_{\ell_1\ell_2\ell_3}$ and $B^{X'Y'Z'}_{\ell_1'\ell_2'\ell_3'}$ are not distinct signals if there exists a permutation $\sigma$ that simultaneously maps $X'Y'Z'$ to $XYZ$ and $\ell_1'\ell_2'\ell_3'$ to $\ell_1\ell_2\ell_3$. It turns out that we can write a sum over distinct signals explicitly as
\begin{gather*}
    \sum_{\text{distinct signals}} = \underbrace{\sum_{\ell_1=\ell_2=\ell_3}\sum_{[XYZ]}}_{\text{sum }1} + \underbrace{\sum_{\ell_1=\ell_2\neq \ell_3}\sum_{[XY]Z}}_{\text{sum }2} + \underbrace{\sum_{\ell_1<\ell_2<\ell_3}\sum_{XYZ}}_{\text{sum }3}.
\end{gather*}
With the same definition again for the $[\cdot]$ notation. For example:
\begin{gather*}
    \{[XY]Z|X,Y,Z\in\{\psi_\kappa, \psi_\gamma\}\} = \{\psi_\kappa\psi_\kappa\psi_\kappa, \psi_\kappa\psi_\gamma\psi_\kappa, \psi_\gamma\psi_\gamma\psi_\kappa,\psi_\kappa\psi_\kappa\psi_\gamma, \psi_\kappa\psi_\gamma\psi_\gamma, \psi_\gamma\psi_\gamma\psi_\gamma\}.
\end{gather*}

It follows to show that the sets that these sums sum over form a partition of the set of all distinct signals. Clearly, all 3 sets are pairwise disjoint (no common elements) because of the criteria for the $l_i$'s. To show that their union covers the set of distinct signals, start by considering an arbitrary signal. Its $l$ configuration will trivially correspond to exactly one of the three sums. If it is sum 1, then we are free to permute the $XYZ$'s by virtue of the $l$'s being identical, so we will be able to match the $XYZ$ configuration to one of the elements of $\{\{XYZ\}\}$. Similarly, if the $l$ configuration corresponds to sum 2, then we are free to permute the $XY$ configuration to match with one of the elements of $\{\{XY\}Z\}$. The $Z$ value does not matter because any $Z$ value is accounted for. For sum 3, we can argue that we can switch around the order of the $l$'s to satisfy $\ell_1 < l2 < l3$ and the corresponding $XYZ$ configuration will be accounted for in sum 3 because all $XYZ$ combinations are counted. Finally, it is simple to verify that no distinct signal is counted more than once within each sum.

To calculate the elements of the covariance matrix, consider the following. Every element of the Fisher matrix can be seen as an inner product weighted by the inverse covariance matrix. We can choose how we order the vectors\footnote{The entries are the derivatives of the bispectra}. We organize the vectors according to the sum of 1, 2, and 3 parts first. Then by $l$ configuration. Within each $l$ configuration, we can choose any ordering for the $XYZ$ configurations. The covariance matrix now becomes a block matrix with each block corresponding to an $l_i$ and $l_i'$ configuration. When wick contracting using the Gaussian approximation, every block matrix where $(\ell_1,\ell_2,\ell_3)\neq (\ell_1',\ell_2',\ell_3')$ vanishes. It can then be shown that the entries of each block matrix are given as
\begin{gather*}
    \br{\Cov_{\ell_1\ell_2\ell_3}}^{XYZ,X'Y'Z'} = \tilde C^{XX'}_{\ell_1}\tilde C^{YY'}_{\ell_2}\tilde C^{ZZ'}_{\ell_3} + \delta_{\ell_1\ell_2}\tilde C^{XY'}_{\ell_1}\tilde C^{YX'}_{\ell_2}\tilde C^{ZZ'}_{\ell_3} +\delta_{\ell_2\ell_3}\tilde C^{XX'}_{\ell_1}\tilde C^{YZ'}_{\ell_2}\tilde C^{ZX'}_{\ell_3} \\ + \delta_{\ell_3\ell_1}\tilde C^{XZ'}_{\ell_1}\tilde C^{YY'}_{\ell_2}\tilde C^{ZX'}_{\ell_3} +\delta_{\ell_1\ell_2}\delta_{\ell_2\ell_3}\br{\tilde C^{XY'}_{\ell_1}\tilde C^{YZ'}_{\ell_2}\tilde C^{ZX'}_{\ell_3} +\tilde C^{XZ'}_{\ell_1}\tilde C^{YX'}_{\ell_2}\tilde C^{ZY'}_{\ell_3}}.
\end{gather*}
With our ordering, this means that the covariance matrix is again a diagonal block matrix, now with blocks of size $4 \times 4$ (sum 1), $6 \times 6$ (sum 2), and $8 \times 8$ (sum 3).

\subsection{Explicit form for inverse covariance matrix}
The Fisher matrix above can be significantly simplified and written as
\begin{equation*}
F_{\alpha\beta} = \sum_{\ell_1 \leq \ell_2 \leq \ell_3} \frac{\mathcal P _{\ell_1\ell_2\ell_3}}{6}
\sum_{XYZ}\sum_{X'Y'Z'} 
\partial_\alpha B^{X Y Z}_{\ell_1 \ell_2 \ell_3} 
(\tilde C^{-1})^{X X'}_{\ell_1}
(\tilde C^{-1})^{Y Y'}_{\ell_2}
(\tilde C^{-1})^{Z Z'}_{\ell_3}
\partial_\beta B^{X' Y' Z'}_{\ell_1 \ell_2 \ell_3}
\end{equation*}
where, in the case of two tracers, 
\begin{equation*}
    C_{l} := \begin{pmatrix}
        \tilde C^{\psi_1\psi_1}_l & \tilde C^{\psi_1\psi_2}_l \\
        \tilde C^{\psi_1\psi_2}_l & \tilde C^{\psi_2\psi_2}_l
    \end{pmatrix}
\end{equation*}
and $\mathcal P_{\ell_1\ell_2\ell_3}$ is defined as the number of distinct permutations that can be made with $\ell_1\ell_2\ell_3$. This form was, for example, used in \cite{Kalaja_2021}\footnote{Note that in \cite{Kalaja_2021} this is based on a previous equation summing over \textit{all} $l_i$ (so including permutations of each configuration) which is missing a factor of $1/6$.}.

To show that the above is the same as the formula for the Fisher matrix given earlier, first partition the sum in the same way and collect all terms that fit in the different categories.
\begin{align*}
    F_{\alpha\beta} &= \sum_{\ell_1=\ell_2=\ell_3} \sum_{[XYZ]}\sum_{[X'Y'Z']} \partial_{\alpha} B^{XYZ}_{\ell_1 \ell_2 \ell_3}\br{\frac{\mathcal P _{\ell_1\ell_2\ell_3}}{6}\sum_{d. p. XYZ} \sum_{d. p. X'Y'Z'}(\tilde C^{-1})^{X X'}_{\ell_1}
    (\tilde C^{-1})^{Y Y'}_{\ell_2}
    (\tilde C^{-1})^{Z Z'}_{\ell_3}}
    \partial_\beta B^{X' Y' Z'}_{\ell_1 \ell_2 \ell_3} \\
    &+ \sum_{\ell_1=\ell_2\neq \ell_3} \sum_{[XY]Z}\sum_{[X'Y']Z'} \partial_{\alpha} B^{XYZ}_{\ell_1 \ell_2 \ell_3}\br{\frac{\mathcal P _{\ell_1\ell_2\ell_3}}{6}\sum_{d. p. XY} \sum_{d. p. X'Y'}(\tilde C^{-1})^{X X'}_{\ell_1}
    (\tilde C^{-1})^{Y Y'}_{\ell_2}
    (\tilde C^{-1})^{Z Z'}_{\ell_3}}
    \partial_\beta B^{X' Y' Z'}_{\ell_1 \ell_2 \ell_3} \\
    &+ \sum_{\ell_1<\ell_2<\ell_3} \sum_{XYZ}\sum_{X'Y'Z'} \partial_{\alpha} B^{XYZ}_{\ell_1 \ell_2 \ell_3}(\tilde C^{-1})^{X X'}_{\ell_1}
    (\tilde C^{-1})^{Y Y'}_{\ell_2}
    (\tilde C^{-1})^{Z Z'}_{\ell_3}
    \partial_\beta B^{X' Y' Z'}_{\ell_1 \ell_2 \ell_3},
\end{align*}
where ``$d. p.$'' stands for ``distinct permutations''.
The entries above are then the entries of the inverses of the block matrices mentioned earlier. This can be checked. For example, for the $\ell_1 = \ell_2 = \ell_3$ sum the multiplication of block matrices corresponding to the same $l_i$ configuration can be written as:
\begin{gather*}
    \sum_{[X'Y'Z']} \br{\frac{\mathcal P _{\ell_1\ell_2\ell_3}}{6} \sum_{d. p. XYZ} \sum_{d. p. X'Y'Z'}(\tilde C^{-1})^{X X'}_{l}
    (\tilde C^{-1})^{Y Y'}_{l}
    (\tilde C^{-1})^{Z Z'}_{l}} \br{\tilde C^{X'X''}_l \tilde C^{Y'Y''}_l \tilde C^{Z'Z''}_l + \text{perms } X''Y''Z''} \\
    = \frac{\mathcal P _{\ell_1\ell_2\ell_3}}{6}\sbr{\br{\sum_{d. p. XYZ} \sum_{X'Y'Z'}(\tilde C^{-1})^{X X'}_{l}
    (\tilde C^{-1})^{Y Y'}_{l}
    (\tilde C^{-1})^{Z Z'}_{l}} \tilde C^{X'X''}_l \tilde C^{Y'Y''}_l \tilde C^{Z'Z''}_l} + \text{perms } X''Y''Z'' \\
    = \frac{\mathcal P _{\ell_1\ell_2\ell_3}}{6} \sum_{d. p. XYZ} \delta_{XX''}\delta_{YY''}\delta_{ZZ''} + \text{perms } X''Y''Z''
    = \frac{\mathcal P _{\ell_1\ell_2\ell_3}}{6} \delta_{[XYZ], [X''Y''Z'']} + \text{perms } X''Y''Z''\\
    = \delta_{[XYZ], [X''Y''Z'']} .
\end{gather*}
The sum over the different wick contractions will similarly cancel with the $\mathcal P_{\ell_1\ell_2\ell_3} / 6$ factor for the $\ell_1 = \ell_2 \neq \ell_3$ sum. For the $\ell_1 < \ell_2 < \ell_3$ sum, the proof is similar as well, except no cancellation is required.

The same type of simplification can be made in the Fisher matrix for the power spectrum, though it doesn't offer any significant benefits compared to our current $3\times 3$ block matrix form.

\subsection{Signal to Noise Ratio (SNR)}
To quantify the detectability of the lensing spectra, we introduce an overall amplitude of our signal, $A$, with fiducial value 1 as an experimental parameter and compute $F_{AA}$. Obviously, $\partial_A\br{A B^{XYZ}_{\ell_1\ell_2\ell_3}}|_{A=1} = B^{XYZ}_{\ell_1\ell_2\ell_3}$, so we find
\begin{equation*}
    \br{\frac{S}{N}}^2 := F_{AA} =
\sum_{XYZ, X'Y'Z'} \sum_{\ell_1\leq \ell_2 \leq \ell_3} \frac{\mathcal P _{\ell_1\ell_2\ell_3}}{6}
B^{X Y Z}_{\ell_1 \ell_2 \ell_3} 
(\tilde C^{-1})^{X X'}_{\ell_1}
(\tilde C^{-1})^{Y Y'}_{\ell_2}
(\tilde C^{-1})^{Z Z'}_{\ell_3}
B^{X' Y' Z'}_{\ell_1 \ell_2 \ell_3}.
\end{equation*}
The equation for the SNR of the power spectra is identical in form.

\subsection{Fisher matrix of power- + bispectra}
To compute the Fisher matrix of an experiment measuring both lensing power and bispectra, we are required to compute and invert the full covariance matrix. If we keep assuming that the measurements are close enough to Gaussian to be able to use Wick contractions to a good approximation, the full covariance matrix simplifies trivially. The correlation between a power and bispectrum estimator contains an odd (5) amount of fields and thus always vanishes. We are thus allowed to simply add the Fisher matrices of the power and bispectra to compute the combined Fisher matrix.  

\section{Shear equals twice spin raised lensing potential}\label{sec:shear}

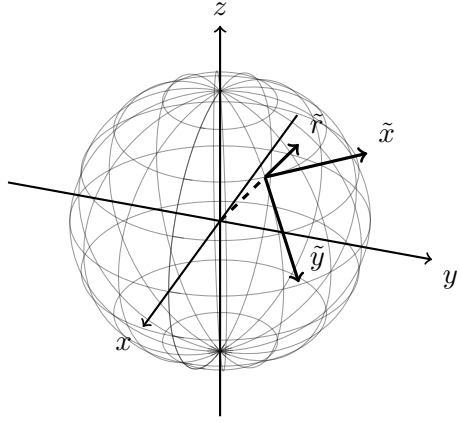
\begin{figure}[!]
    \centering
    \begin{tikzpicture}    
    \tdplotsetmaincoords{60}{110}
    \begin{scope}[tdplot_main_coords]

            \draw[thick, ->] (-3,0,0) -- (3,0,0) node[anchor=north east] {$x$};
            \draw[thick, ->] (0,-3,0) -- (0,3,0) node[anchor=north west] {$y$};
            \draw[thick, ->] (0,0,-3) -- (0,0,3) node[anchor=south] {$z$};

            \draw[very thick, dashed] (0, 0, 0) -- ({2*sin(45)*cos(45)},{2*sin(45)*sin(45)},{2*cos(45)});
            \draw[very thick, ->] ({2*sin(45)*cos(45)},{2*sin(45)*sin(45)},{2*cos(45)}) -- ({3.5*sin(45)*cos(45)},{3.5*sin(45)*sin(45)},{3.5*cos(45)}) node[anchor=south west] {$\tilde r$};
            \draw[very thick, ->] ({2*sin(45)*cos(45)},{2*sin(45)*sin(45)},{2*cos(45)}) -- ({2*sin(45)*cos(45) + 1.5*cos(45)*cos(45)},{2*sin(45)*sin(45) + 1.5*cos(45)*sin(45)},{2*cos(45) - 1.5*sin(45)}) node[anchor=south west] {$\tilde y$};
            \draw[very thick, ->] ({2*sin(45)*cos(45)},{2*sin(45)*sin(45)},{2*cos(45)}) -- ({2*sin(45)*cos(45) - 1.5*sin(45)},{2*sin(45)*sin(45) +1.5*cos(45)},{2*cos(45)}) node[anchor=south west] {$\tilde x$};
        \foreach \u in {0,22.5,...,180} { 
        \foreach \v in {0,5,...,360} { 
            \pgfmathsetmacro\x{2*sin(\u)*cos(\v)}
            \pgfmathsetmacro\y{2*sin(\u)*sin(\v)}
            \pgfmathsetmacro\z{2*cos(\u)}

            \pgfmathsetmacro\lx{2*sin(\u)*cos(\v+5)}
            \pgfmathsetmacro\ly{2*sin(\u)*sin(\v+5)}
            \pgfmathsetmacro\lz{2*cos(\u)}

            \draw[opacity=0.4] (\x,\y,\z) -- (\lx,\ly,\lz);
        }
    }

    \foreach \v in {0,22.5,...,360} { 
        \foreach \u in {0,5,...,180} { 
            \pgfmathsetmacro\x{2*sin(\u)*cos(\v)}
            \pgfmathsetmacro\y{2*sin(\u)*sin(\v)}
            \pgfmathsetmacro\z{2*cos(\u)}

            \pgfmathsetmacro\nx{2*sin(\u+5)*cos(\v)}
            \pgfmathsetmacro\ny{2*sin(\u+5)*sin(\v)}
            \pgfmathsetmacro\nz{2*cos(\u+5)}

            \draw[opacity=0.4] (\x,\y,\z) -- (\nx,\ny,\nz);
        }
    }

    \end{scope}
\end{tikzpicture}    
    \caption{
        $(\tilde r, \tilde y, \tilde x)$ coordinates defined for a point on the unit sphere. These act as ordinary cartesian coordinates but rotated such that, at the associated point on $S^2$, $\hat{\tilde r}$ points straight out of the unit sphere, $\hat{\tilde y}$ is parallel to the great arc with constant $\phi$ and $\hat{\tilde x}$ is parallel to the great arc with constant $\theta$. These coordinates are used to define the shear and convergence in terms of the lensing potential.
        }
    \label{fig:tildecoords}
\end{figure}

Consider a point on $S^2$, $(r_0, \theta_0, \phi_0)$, at which we observe some cosmological object. We can then define a set of cartesian coordinates $(\tilde r, \tilde y, \tilde x)$ as shown in figure \ref{fig:tildecoords}.

Note that it is not obvious whether to define these coordinates at the point where the lensed light hits $S^2$ or the unlensed light hits $S^2$. We will assume that lensing effects are sufficiently weak that either definition works. We can then express the tilde coordinates in terms of spherical coordinates either by applying a rotation matrix or by calculating the $r, \theta, \phi$ derivatives of $(x,y,z)$ coordinates at $(r_0, \theta_0, \phi_0)$ to find $\hat{\tilde{\mathbf r}}$, $\hat{\tilde{\boldsymbol{\theta}}}$, and $\hat{\tilde{\boldsymbol{\phi}}}$ and then take inner products. Regardless, we find
\begin{align}
    \tilde r &= r\sin\theta\cos\phi\sin\theta_0\cos\phi_0+r\sin\theta\sin\phi\sin\theta_0\sin\phi_0+r\cos\theta\cos\theta_0,\\
    \tilde y &= r\sin\theta\cos\phi\cos\theta_0\cos\phi_0+r\sin\theta\sin\phi\cos\theta_0\sin\phi_0-r\cos\theta\sin\theta_0,\\
    \tilde x &= -r\sin\theta\cos\phi\sin\phi_0+r\sin\theta\sin\phi\cos\phi_0.
\end{align}
This gives derivatives
\begin{align*}
    \frac{\partial}{\partial \theta} = & \left( r \cos \theta \cos \phi \sin \theta_0 \cos \phi_0 + r \cos \theta \sin \phi \sin \theta_0 \sin \phi_0 - r \sin \theta \cos \theta_0 \right) \frac{\partial}{\partial \tilde{r}} \\
    & + \left( r \cos \theta \cos \phi \cos \theta_0 \cos \phi_0 + r \cos \theta \sin \phi \cos \theta_0 \sin \phi_0 + r \sin \theta \sin \theta_0 \right) \frac{\partial}{\partial \tilde{y}} \\
    & + \left( -r \cos \theta \cos \phi \sin \phi_0 + r \cos \theta \sin \phi \cos \phi_0 \right) \frac{\partial}{\partial \tilde{x}}.\\
    \frac{\partial}{\partial \phi} = & \left( -r \sin \theta \sin \phi \sin \theta_0 \cos \phi_0 + r \sin \theta \cos \phi \sin \theta_0 \sin \phi_0 \right) \frac{\partial}{\partial \tilde{r}} \\
& + \left( -r \sin \theta \sin \phi \cos \theta_0 \cos \phi_0 + r \sin \theta \cos \phi \cos \theta_0 \sin \phi_0 \right) \frac{\partial}{\partial \tilde{y}} \\
& + \left( r \sin \theta \sin \phi \sin \phi_0 + r \sin \theta \cos \phi \cos \phi_0 \right) \frac{\partial}{\partial \tilde{x}}.
\end{align*}
Evaluated at our point of interest we obtain $\partial_\theta=\partial_{\tilde y}$ and $\partial_\phi=\sin\theta_0\partial_{\tilde x}$. The second-order derivatives can then be obtained using the first-order derivatives. We can immediately evaluate them at the point to get
\begin{align*}
    \partial_\phi^2|_{(r_0,\theta_0,\phi_0)}&=-\sin^2\theta_0\partial_{\tilde r}-\sin\theta_0\cos\theta_0\partial_{\tilde y}+\sin^2\theta_0\partial^2_{\tilde x},\\
    \partial_\theta\partial_\phi|_{(r_0,\theta_0,\phi_0)}&=\cos\theta_0\partial_{\tilde x}+\sin\theta_0\partial_{\tilde x}\partial_{\tilde y},\\
    \partial_\theta^2|_{(r_0,\theta_0,\phi_0)}&=-\partial_{\tilde r}+\partial^2_{\tilde y}.
\end{align*}
Thus, at $(r_0, \theta_0, \phi_0)$,
\begin{gather*}
    \frac{1}{2}\eth_1(\eth_0\psi) = \frac{1}{2}\sin\theta(\partial_\theta+\frac{i}{\sin\theta}\partial_\phi)(\frac{1}{\sin\theta}(\partial_\theta+\frac{1}{\sin\theta}\partial_\phi))\\
    =\frac{\partial^2 \psi}{\partial \theta^2} - \frac{\cos\theta}{\sin\theta} \frac{\partial \psi}{\partial \theta} + \frac{2 i}{\sin\theta} \frac{\partial^2 \psi}{\partial \theta \partial \phi} - \frac{1}{\sin^2\theta} \frac{\partial^2 \psi}{\partial \phi^2} - 2 i \frac{\cos\theta}{\sin^2\theta} \frac{\partial \psi}{\partial \phi}\\
    =\frac{1}{2}(\partial_{\tilde y}^2 - \partial_{\tilde x}^2 + 2i\partial_{\tilde x}\partial_{\tilde y})\psi
    =\gamma_1 + i\gamma_2 = \gamma.
\end{gather*}

\section{Numerical derivative}
\label{sec:derivatives}
The derivatives with respect to cosmological parameters were taken with a central difference formula, i.e.
\begin{equation*}
    f'(x) = \frac{f(x+h) - f(x-h)}{2h} + \mathcal O (h^2).
\end{equation*}
Each change of the cosmological parameters requires a recalculation of the entire cosmology, making it computationally expensive. For the approximation to be accurate, a balance needs to be found between numerical errors for small $h$ and a larger $\mathcal O(h^2)$ error for larger $h$. The $h$ values chosen for each parameter are shown in table \ref{tab:cosmo-params-diff} and are similar to the values used in \cite{Namikawa_2016}\footnote{As confirmed during a conversation with the author.}.

\begin{table}[!]
    \centering
    \begin{tabular}{|l|c|c|}
    \hline
    Parameter & Fiducial value & Finite difference ($h$) \\
    \hline
    $H$            & $67.4$               & fiducial\,$\times0.1$   \\
    $\Omega_b h^2$ & $0.0223$             & fiducial\,$\times0.1$   \\
    $\Omega_c h^2$ & $0.119$              & fiducial\,$\times0.005$ \\
    $n_s$          & $0.965$              & fiducial\,$\times0.005$ \\
    $A_s$          & $2.13\times10^{-9}$  & fiducial\,$\times0.1$   \\
    $\tau$         & $0.063$              & fiducial\,$\times 0.1$   \\
    $m_\nu$        & $0.06$               & fiducial\,$\times0.1$   \\
    $w_0$          & $-1$                 & $0.03$                  \\
    \hline
    \end{tabular}
    \caption{Fiducial cosmological parameters and their finite-difference steps}
    \label{tab:cosmo-params-diff}
\end{table}

To test the accuracy, we varied $h$ by $\pm 5\%$ and $\pm 10\%$ and compared the relative change in the derivative of the equilateral lensing bispectra and the lensing power spectra. As long as numerical noise doesn't dominate, it can be shown that the relative error in our approximation is approximately 5 times the relative difference that taking $h \rightarrow h(1 \pm 0.1)$ leads to. We thus conclude that, based on the tests conducted, the derivatives are almost always computed with up to one percent error. The exception is the derivative with respect to neutrino masses, which introduces a larger error of around 10 percent.

\pagebreak
\section{Additional plots}
\label{sec:lambdacdmconstraints}

\begin{figure}[!]
    \centering
    \includegraphics[width=\textwidth]{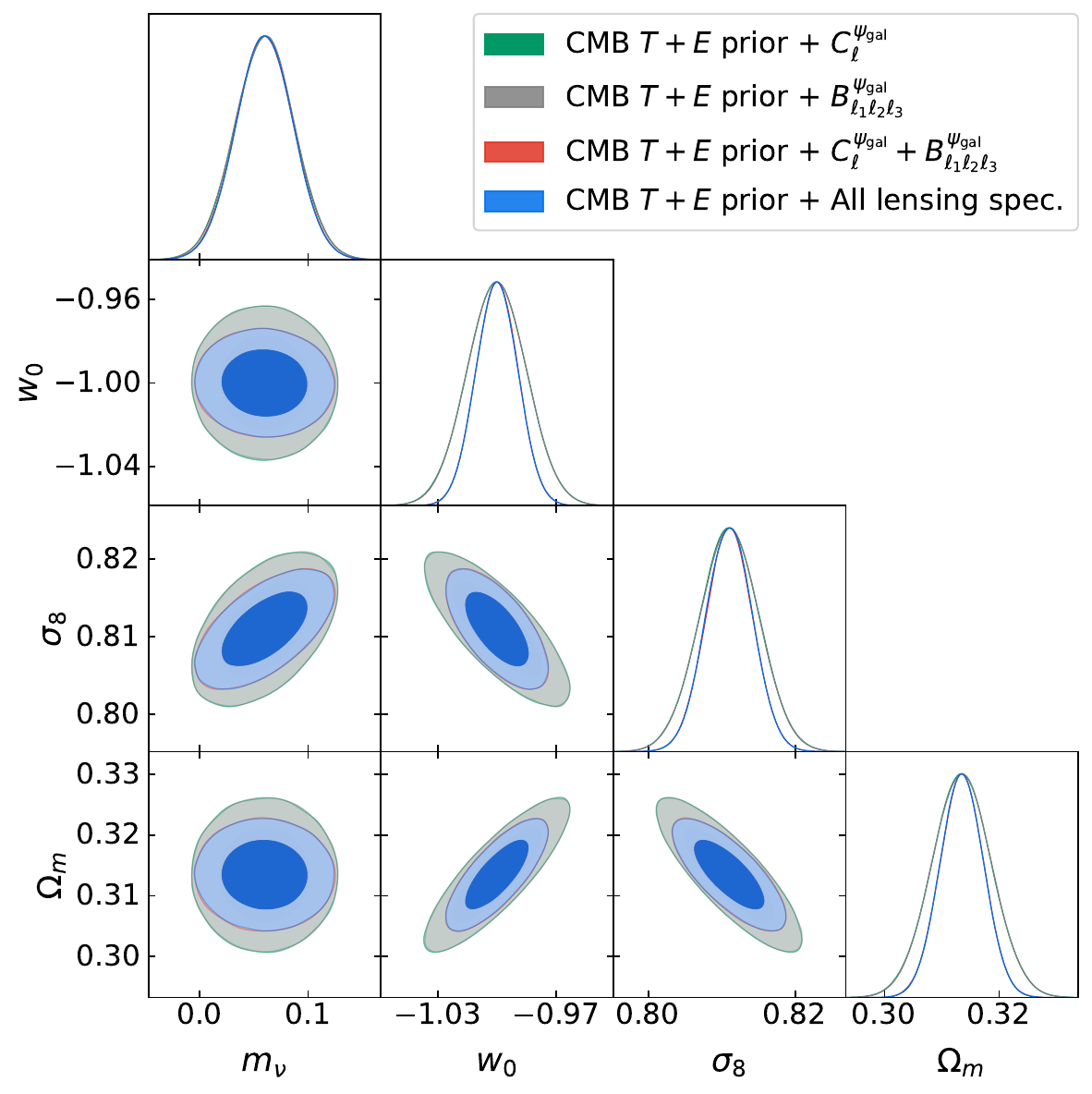}
    \caption{Same as figures \ref{fig:paramconstraintstightcmb} and \ref{fig:paramconstraintstightgal}, except for different combinations of data.}
    \label{fig:paramconstraintslpsprior}
\end{figure}

\begin{figure}[!]
    \centering
    \includegraphics[width=\textwidth]{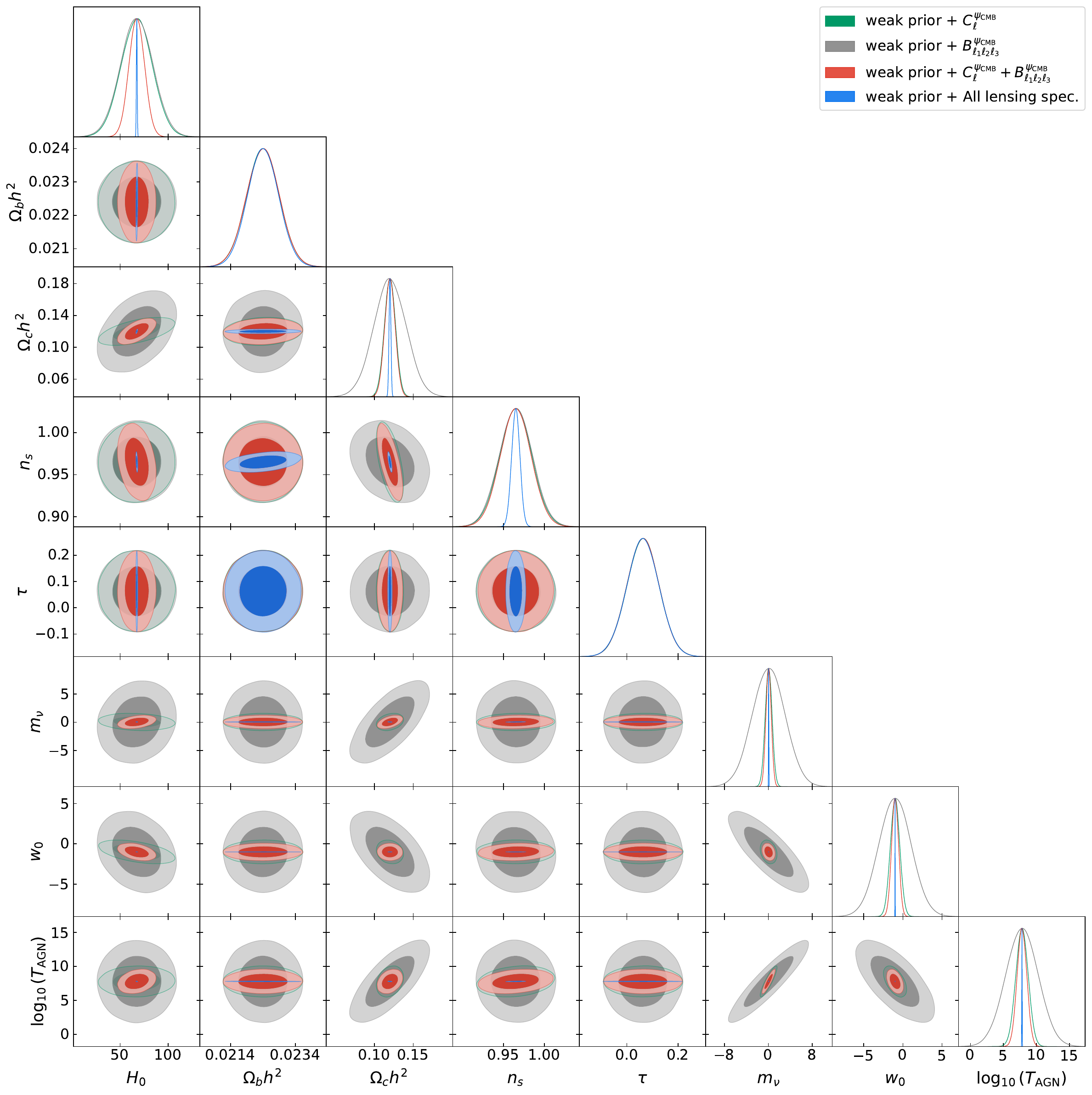}
    \caption{Same as figure \ref{fig:paramconstraintstightcmb}, except for the standard $\Lambda$CDM parameters.}
    \label{fig:paramconstraintsallcmb}
\end{figure}

\begin{figure}[!]
    \centering
    \includegraphics[width=\textwidth]{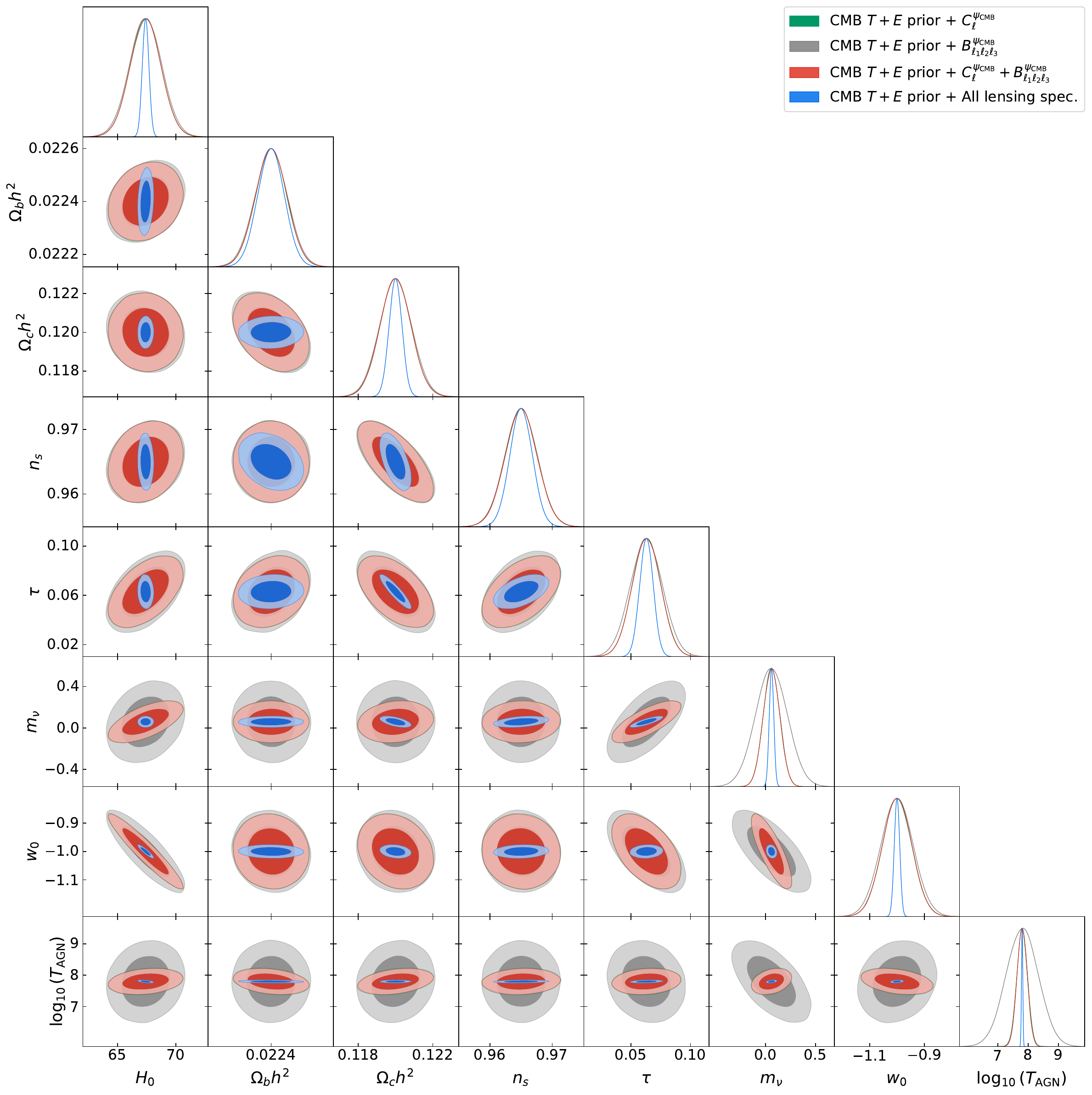}
    \caption{Same as figure \ref{fig:paramconstraintstightcmbcmbprior}, except for the standard $\Lambda$CDM parameters.}
    \label{fig:paramconstraintsallcmbcmbprior}
\end{figure}

\begin{figure}[!]
    \centering
    \includegraphics[width=\textwidth]{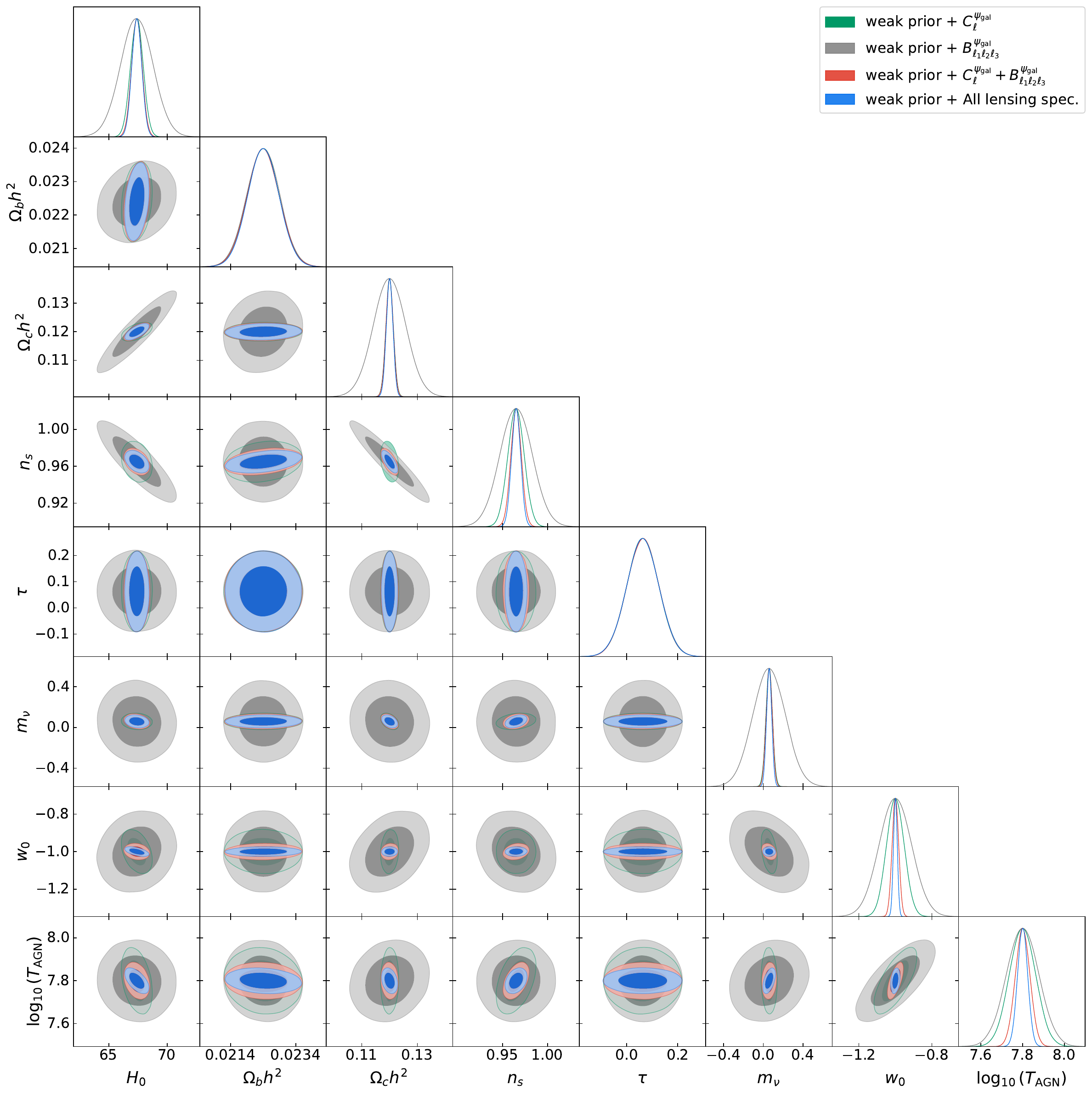}
    \caption{Same as figure \ref{fig:paramconstraintstightgal}, except for the standard $\Lambda$CDM parameters.}
    \label{fig:paramconstraintsallgal}
\end{figure}

\begin{figure}[!]
    \centering
    \includegraphics[width=\textwidth]{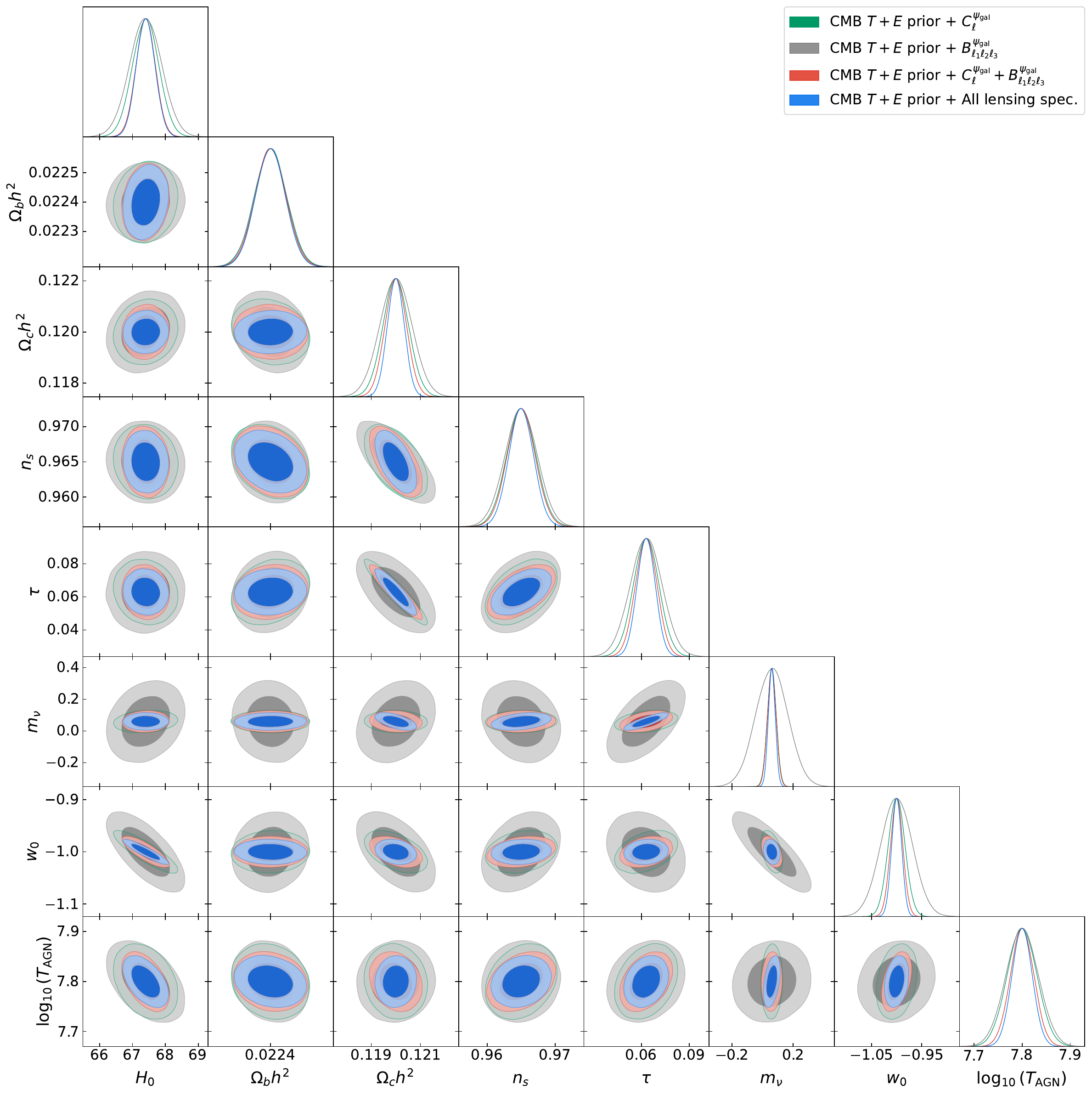}
    \caption{Same as figure \ref{fig:paramconstraintstightgalcmbprior}, except for the standard $\Lambda$CDM parameters.}
    \label{fig:paramconstraintsallgalcmbprior}
\end{figure}

We do not expect to use weak lensing surveys to competitively constrain the main $\Lambda$CDM parameters. Despite this, we include these constraints for completeness in figures \ref{fig:paramconstraintsallcmb}, \ref{fig:paramconstraintsallcmbcmbprior}, \ref{fig:paramconstraintsallgal}, \ref{fig:paramconstraintsallgalcmbprior}, and in table \ref{tab:paramconstraintsall}. All survey parameters are the same as in section \ref{sec:results}. We also include a plot (figure \ref{fig:paramconstraintslpsprior}) similar to those found in section \ref{sec:results} but comparing (for weak priors) the constraints from all powerspectra, all powerspectra + galaxy bispectra, all powerspectra + CMB bispectra, and all powerspectra + all bispectra. This makes it clearer which bispectra add the most information. We see there clearly that if we already use lensing powerspectra from CMB and galaxies, the CMB bispectrum adds little information, while the galaxy lensing bispectrum measurably improves constraints.
\begin{table}
\centering
\scriptsize
\begin{tabular}{|l|r|rrr|rrr|rrr|}
\hline
\multicolumn{11}{|c|}{weak priors (with and without post-Born corrections)} \\
\hline
&& \multicolumn{3}{c|}{CMB lensing} & \multicolumn{3}{c|}{Gal. lensing} & \multicolumn{3}{c|}{CMB $\times$ Gal. lensing} \\
\hline
Par & prior & $C_\ell$ & $B$ & $C_\ell+B$ & $C_\ell$ & $B$ & $C_\ell+B$ & $C_\ell$ & $B$ & $C_\ell+B$ \\
\hline
$H_0$ & 17 & 16 & 17 & 8.1 (2.0) & 0.5 & 1.4 & 0.45 (1.2) & 0.52 & 1.4 & 0.42 (1.2)\\
\rowcolor{blue!10}
 & & & 16 & 7.3 (2.2) & & 1.4 & 0.45 (1.2) & & 0.8 & 0.40 (1.3) \\

$10^4\Omega_b h^2$ & 5 & 5.0 & 5.0 & 5.0 (1.0) & 4.9 & 5.0 & 4.9 (1.0) & 4.8 & 5.0 & 4.8 (1.0) \\
\rowcolor{blue!10}
 & & & 5.0 & 5.0 (1.0) & & 5.0 & 4.9 (1.0) & & 4.9 & 4.7 (1.0) \\

$10^3\Omega_c h^2$ & 290 & 7.1 & 21 & 6.7 (1.1) & 1.3 & 5.9 & 1.3 (1.0) & 1.2 & 5.6 & 1.2 (1.0)\\
\rowcolor{blue!10}
 & & & 22 & 5.0 (1.4) & & 5.8 & 1.3 (1.0) & & 2.6 & 1.1 (1.1) \\

$10^3(n_s-1)$ & 20 & 19 & 20 & 19 (1.0) & 9.1 & 18 & 5.9 (1.5) & 5.5 & 18 & 5.0 (1.1) \\
\rowcolor{blue!10}
 & & & 20 & 18 (1.1) & & 18 & 6.0 (1.5) & & 10 & 4.9 (1.1) \\

$10^2\tau$ & 6 & 6 & 6 & 6 (1.0) & 6 & 6 & 6 (1.0) & 6 & 6 & 6 (1.0)\\
\rowcolor{blue!10}
 & & & 6 & 6 (1.0) & & 6 & 6 (1.0) & & 6 & 6 (1.0) \\

$m_\nu$ & 10 & 0.6 & 3.0 & 0.49 (1.3) & 0.03 & 0.16 & 0.03 (1.1) & 0.03 & 0.15 & 0.03 (1.0)\\
\rowcolor{blue!10}
 & & & 2.2 & 0.50 (1.2) & & 0.17 & 0.03 (1.1) & & 0.15 & 0.03 (1.1) \\

$w_0$ & 10 & 0.6 & 2.1 & 0.44 (1.4) & 0.05 & 0.09 & 0.02 (2.7) & 0.02 & 0.05 & 0.01 (1.4)\\
\rowcolor{blue!10}
 & & & 1.0 & 0.37 (1.6) & & 0.09 & 0.02 (2.7) & & 0.05 & 0.01 (1.4) \\

$\log_{10}(T_{\rm AGN})$ & 10 & 0.9 & 2.5 & 0.74 (1.3) & 0.06 & 0.08 & 0.04 (1.7) & 0.03 & 0.06 & 0.02 (1.3)\\
\rowcolor{blue!10}
 & & & 1.9 & 0.75 (1.2) & & 0.08 & 0.04 (1.7) & & 0.06 & 0.02 (1.3) \\

\hline
\multicolumn{11}{|c|}{CMB $T+E$ priors (with and without post-Born corrections)}\\
\hline
&& \multicolumn{3}{c|}{CMB lensing} & \multicolumn{3}{c|}{Gal. lensing} & \multicolumn{3}{c|}{CMB $\times$ Gal. lensing}\\
\hline
Par & prior & $C_\ell$ & $B$ & $C_\ell+B$ & $C_\ell$ & $B$ & $C_\ell+B$ & $C_\ell$ & $B$ & $C_\ell+B$\\
\hline
$H_0$ & 1.4 & 1.3 & 1.4 & 1.3 (1.0) & 0.4 & 0.5 & 0.29 (1.4) & 0.39 & 0.48 & 0.28 (1.4)\\
\rowcolor{blue!10}
 & & & 1.4 & 1.3 (1.0) & & 0.5 & 0.29 (1.4) & & 0.48 & 0.28 (1.4) \\

$10^4\Omega_b h^2$ & 0.65 & 0.61 & 0.63 & 0.61 (1.0) & 0.57 & 0.56 & 0.54 (1.1) & 0.54 & 0.56 & 0.52 (1.0)\\
\rowcolor{blue!10}
 & & & 0.64 & 0.61 (1.0) & & 0.56 & 0.54 (1.1) & & 0.56 & 0.52 (1.0) \\

$10^3\Omega_c h^2$ & 0.87 & 0.83 & 0.86 & 0.83 (1.0) & 0.52 & 0.65 & 0.44 (1.2) & 0.38 & 0.62 & 0.35 (1.1)\\
\rowcolor{blue!10}
 & & & 0.85 & 0.82 (1.0) & & 0.65 & 0.44 (1.2) & & 0.63 & 0.34 (1.1) \\

$10^3(n_s-1)$ & 2.6 & 2.6 & 2.6 & 2.6 (1.0) & 2.2 & 2.4 & 2.1 (1.0) & 1.9 & 2.3 & 1.8 (1.1)\\
\rowcolor{blue!10}
 & & & 2.6 & 2.6 (1.0) & & 2.4 & 2.1 (1.0) & & 2.3 & 1.8 (1.1) \\

$10^2\tau$ & 2.1 & 1.2 & 1.4 & 1.2 (1.0) & 0.8 & 1.0 & 0.68 (1.2) & 0.61 & 1.0 & 0.57 (1.1)\\
\rowcolor{blue!10}
 & & & 1.3 & 1.2 (1.0) & & 1.0 & 0.68 (1.2) & & 1.0 & 0.57 (1.1) \\

$m_\nu$ & 0.22 & 0.08 & 0.16 & 0.08 (1.0) & 0.03 & 0.11 & 0.03 (1.0) & 0.02 & 0.09 & 0.02 (1.0)\\
\rowcolor{blue!10}
 & & & 0.16 & 0.08 (1.0) & & 0.11 & 0.03 (1.0) & & 0.10 & 0.02 (1.0) \\

$w_0$ & 0.06 & 0.05 & 0.06 & 0.05 (1.0) & 0.02 & 0.03 & 0.01 (1.4) & 0.01 & 0.03 & 0.01 (1.4)\\
\rowcolor{blue!10}
 & & & 0.06 & 0.05 (1.0) & & 0.03 & 0.01 (1.4) & & 0.03 & 0.01 (1.4) \\

$\log_{10}(T_{\rm AGN})$ & 16 & 0.18 & 0.54 & 0.17 (1.0) & 0.03 & 0.03 & 0.02 (1.3) & 0.03 & 0.03 & 0.02 (1.3)\\
\rowcolor{blue!10}
 & & & 0.48 & 0.17 (1.0) & & 0.03 & 0.02 (1.3) & & 0.03 & 0.02 (1.3) \\

\hline
\end{tabular}
        \caption{Same as table \ref{tab:paramconstraintstight}, except for $\Lambda$CDM parameters.}
        \label{tab:paramconstraintsall}
\end{table}

\end{document}